\documentclass[iop,numberedappendix] {emulateapj}

\usepackage{graphicx}  
\usepackage{dcolumn}   
\usepackage{bm}        
\usepackage{amssymb}   
\usepackage{hyperref}
\usepackage{amsmath}
\usepackage{color}
\usepackage{multirow}
\usepackage{color}

\begin{document}

\newcommand{\meta}{\boldsymbol\eta}
\newcommand{\bn}{\boldsymbol\nabla}
\newcommand{\p}{\partial}

\title{Evolving waves and turbulence in the outer corona and inner heliosphere: the accelerating expanding box.
}

\author{Anna Tenerani} \author{Marco Velli}  
\affiliation{EPSS, UCLA, Los Angeles, CA}

\begin{abstract}

 Alfv\'enic fluctuations in the solar wind display many properties reflecting an ongoing nonlinear cascade, e.g. a well-defined spectrum in frequency, together with some characteristics more commonly associated with the linear propagation  of waves from the Sun, such as the variation of fluctuation amplitude with distance, dominated by solar wind expansion effects. Therefore both nonlinearities and expansion must be included simultaneously in any successful model of solar wind turbulence evolution. Because of the disparate spatial scales involved, direct numerical simulations of turbulence in the solar wind represent an arduous task, especially if one wants to go beyond the incompressible approximation. Indeed, most simulations  neglect solar wind expansion effects entirely. Here we develop a numerical model to simulate turbulent fluctuations from the outer corona to 1 AU and beyond, including the sub-Alfv\'enic corona. The accelerating expanding box  (AEB) extends the validity of previous expanding box models  by taking into account both the acceleration of the solar wind and the inhomogeneity of background density and magnetic field.   Our method  incorporates a background accelerating wind within a magnetic field that naturally follows the Parker spiral evolution using a two-scale analysis in which the macroscopic spatial effect coupling fluctuations with background gradients becomes a time-dependent coupling term in a homogeneous box. In this paper we describe the AEB model in detail and discuss its main properties, illustrating its validity by studying  Alfv\'en wave propagation across the Alfv\'en critical point. 
 
\end{abstract}

\maketitle

\section{Introduction}
In this paper we propose and discuss the accelerating expanding box (AEB). The AEB is a three-dimensional numerical model based on the compressible Magnetohydrodynamic (MHD) description of the plasma, that allows numerical studies of the evolution of turbulence and structures in the accelerating, expanding solar wind. The model presented here generalizes the expanding box approximation~\citep{grappin_PRL_1993} to regions close to the sun, overcoming the limitations of a previous version of the model introduced in~\cite{tenerani_JGR_2013} to guarantee the exact conservation of the proper {non-WKB} invariants for counter-propagating Alfv\'en waves {in the limit of small amplitude}~\citep{heinemann_JGR_1980}.

Understanding how the wind expansion affects plasma dynamics and the evolution of fluctuations is the starting point to understand the more general problem of solar wind heating and acceleration via turbulent dissipation and Reynolds stresses (wave-driven winds).  In-situ measurements show that the solar wind is indeed in a turbulent state: while in the slow wind the magnetic energy spectrum displays in general a power-law index close to the standard $-5/3$ predicted by the Kolmogorov phenomenology, hot fast streams originating from coronal holes appear to be permeated  by almost incompressible, large amplitude Alfv\'enic fluctuations~\citep{coleman_PRL_1966, belcher_JGR_1971} characterized by a ``younger" (flatter) spectrum which evolves with heliocentric distance~\citep{bavassano_1982}. Despite the wealth of observations dating back from the '60s, and presently covering distances from 0.3 out to many AU over a wide range of heliographic latitudes (see for instance \cite{horbury_PPCF_2005} and \cite{bruno_2013}), the mechanisms leading to this Alfv\'enic turbulent state and the possible role of such large amplitude fluctuations in the acceleration and heating, specifically of fast streams~\citep{cranmer_2015}, remain one of the main questions concerning the solar wind. In addition, there is no present comprehensive understanding of the variation of turbulence characteristics with solar wind state (slow, fast, et.c.) throughout the solar activity cycle. In particular, a high degree of Alfv\'enicity has been found also in specific slow wind periods, most probably associated with streams originating close to the boundary of coronal holes~\citep{raffa}. 

Alfv\'enic turbulence in fast streams displays velocity-magnetic field correlations corresponding to waves largely propagating away from the sun. The Alfv\'enic  spectrum dominates  the frequency range $f\simeq10 ^{-4}-10^{-2}$~Hz and shows a low frequency part where the fluctuation energy density scales as $E(f)\sim f^{-1}$, and a high frequency one where $E(f)\sim f^{-5/3}$.  Observations indicate that the spectrum evolves towards the same ``standard" state found in the slow wind at increasing heliocentric distances, as the knee separating the steeper from the flatter parts of the spectrum gradually shifts towards the lower frequencies. This is evidence  that a nonlinear energy cascade must be at play, {as was originally suggested by~\cite{coleman_1968}}. At the same time the imbalance between forward and backward propagating perturbations (cross helicity)  is observed to decrease with heliocentric distance~{\citep{roberts}}. This observation is at odds with theory and simulations of homogeneous, incompressible turbulence showing that an initial imbalance between counter-propagating waves would be reinforced by nonlinear interactions via a process called dynamical alignment~\citep{dobro80}. 

In this respect, neither phenomenological theories nor direct numerical simulations of homogeneous turbulence seem adequate to understand the evolution and  energy cascade of Alfv\'enic fluctuations in the solar wind. The major difficulty comes from the fact that the solar wind is an inhomogeneous medium,  complicating both theoretical and  numerical approaches to the problem. The solar wind inhomogeneity comprises many different aspects. 

{First and foremost the overall expansion of the underlying solar atmosphere  affects the evolution of fluctuations through the gradients in the average fields that introduce reflection and coupling among different wave modes, especially at  frequencies low with respect to the expansion rate~\citep{heinemann_JGR_1980,lou_JGR_1993a,lou_JGR_1993b,lou_JGR_1993c,zhou_1989,velli_PRL_1989,velli_91,matthaeus_ApJ_1999}. These couplings exist in the equations quite independently of any normal mode analysis and have an important effect on the energy density in the fluctuations, causing  a slow radial evolution.  This, in linear theory and in the absence of coupling to compressible modes and transverse gradients,  follows the conservation law of wave action~\citep{bretherton_1969, heinemann_JGR_1980}. }
{Not so surprisingly an  overall radial decrease  of the root-mean-square energies $<\delta b^2>$ is observed in the low frequency part of the Alfv\'enic spectrum,  in good agreement with the theory of wave propagation in the expanding solar wind that, for frequencies  larger than the expansion rate, predicts $<\delta b^2>\sim r^{-3}$~\citep{bavassano_1982,roberts_1990}. Nevertheless, this approach based on noninteracting waves fails in accounting for  many other features of the solar wind, such as the observed radial decrease of cross helicity at high frequencies~\citep{velli_93} or the evolution of the Alfv\'en ratio (ratio between the kinetic and  magnetic energy of the fluctuations) with heliocentric distance.

Other aspects of solar wind inhomogeneity might therefore play an important role. Among these are the transverse gradients (shear) between fast and slow solar wind streams, a gradient which also steepens with distance from the Sun evolving into corotating interaction regions. The interaction of fluctuations with such shears was proposed to provide a mechanism for driving the turbulent cascade, even when the solar wind is initially dominated by outwards-type fluctuations~\citep{roberts_1990} and the stream shear interface is stable. Also, results from incompressible nonlinear simulations~\citep{roberts_1992} and transport turbulence models~\citep{zank_1996, matthaeus_1999_PRL,Breech_2005} are in good agreement with observations in the outer heliosphere pointing to the fact that the regions of strong shears found in the ecliptic plane display fluctuations evolving from an almost purely Alfv\'enic state towards a state with small cross-helicity (``mixed" state) more rapidly than in the polar regions, where the flow shear is weaker. Note that \citep{grappin_JGR_1996} also pointed to the importance of such effects in their expanding box simulations with stream shears. However, as shown in ~\citep{vellietal_90, velli_93} the analytical closure models face important difficulties when approaching the Alfv\'en point.}

It becomes clear that any study of wave and turbulence evolution in the solar wind needs to include {nonlinearities,  expansion effects and stream structures}. Unfortunately,  fully nonlinear  eulerian simulations spanning a wide range of spatial scales which could include gradients of the background all the way down to the smallest scales reached through the turbulent cascade (say,  from one AU all the way down to $10^{-4}-{10^{-6}}$~AU),  are still prohibitively expensive in terms of numerical costs, especially if one wants to go beyond the incompressible approximation. 

{Turbulence driven by Alfv\'en wave reflection triggered by the strong inhomogeneities in the lower corona has been investigated numerically, although in most cases using simplified nonlinear models to describe the fluctuations.} Shell model calculations, which approximate incompressible nonlinear interactions in directions orthogonal to the radial to form a set of coupled first order nonlinear PDEs for each perpendicular scale, have shed some light on the evolution of Alfv\'en waves propagating away from the Sun (\citet{verdini_2012} and references therein). For example, \citet{verdini_2012} found that the double power-law spectrum, $-1$ and $-5/3$,  measured in situ might arise from the combination of strong and weak cascades respectively for reflected (inward propagating) and outward modes at the base of the solar corona. {Semi-phenomenological models which adopt ad-hoc terms for turbulent spectral transfer and heating~\citep{Hossain_1995} have also been useful to investigate the role of reflection-driven turbulence in coronal heating and solar wind acceleration~\citep{matthaeus_ApJ_1999, Dmitruk_2001, verdini_2010}.  

Recently, direct} three-dimensional simulations of turbulence in a given solar wind flux tube {extending up to the Alfv\'en critical point (i.e. where the solar wind speed equals the Alfv\'en speed) have been carried out by~\cite{perez_ApJ_2013},  yet  within the reduced MHD framework}. These simulations showed that the anomalous inwards Alfv\'en modes  induced by the inhomogeneity trigger a nonlinear cascade in the sub-Alfv\'enic region via wave-reflection-driven turbulence~\citep{velli_PRL_1989}, eventually developing a $k_\bot^{-1}$ spectrum in wave-number space. The simplifying assumption of incompressibility, an assumption commonly accepted in the studies of Alfv\'enic turbulence, may be justified on the basis of observational evidence: incompressible fluctuations seem to be the only ones to survive at large heliocentric distances, at least in the fast streams. Slow modes, which tend naturally to steepen into shocks, are expected to be rapidly dissipated in the chromosphere and transition region,   while  oblique fast modes  should suffer strong refraction because of the large gradients in those regions. Nevertheless,  compressible effects may arise as the outcome of transient processes which should not be neglected, an interesting example being the parametric decay of large amplitude Alfv\'en waves that may take place in the outer corona and inner heliosphere~\citep{malara_96,DelZanna_AA_2001, tenerani_JGR_2013} {and nonlinear mode conversion from Alfv\'en to slow mode that may contribute to coronal heating and solar wind acceleration via shock formation and dissipation~\citep{suzuki_2005,matsumoto_2012}.}

The expanding box model (EBM) was introduced a few years ago for the purpose of taking into account in a simple way the basic effects of the expansion on a compressible MHD plasma, in the limit of a radially uniform and super-Alfv\'enic solar wind~\citep{velli_92}. The expanding box approximation is based on a separation of spatial scales which allows for a simultaneous description of the large scale evolution due to the expansion of the solar wind and of the more rapid dynamics at smaller scales via a semi-lagrangian approach. {Expansion effects can be taken into account in the traditional EBM easily, by rescaling  variables according to the radial expansion and by including first order effects of the solar wind velocity gradients, provided the dimensions of the box are small with respect to the average heliocentric distance so that a local cartesian geometry is still appropriate~\citep{grappin_PRL_1993,dz_2015}.} The EBM has proved a useful tool, and it has been extensively employed to investigate MHD instabilities and turbulence~\citep{grappin_PRL_1993,rappazzo_2005,dong,verdini_2016}, {even within stream interaction regions~\citep{grappin_JGR_1996},} as well as the onset and development of kinetic instabilities driven by solar wind expansion with the hybrid expanding box~\citep{liewer_JGR_2001,matteini_2006,petr_2013,petr_2015, maneva_2015}. 

The AEB extends and improves the previous EBM to the regions close to the Sun, at a few solar radii, where it becomes necessary not only to include {the radial acceleration of the} solar wind speed directly, but also to extend the validity of the EBM close to the sonic and Alfv\'en critical points by taking into account explicitly the inhomogeneity of the background density and magnetic field. {This model and its derivation therefore explicitly rely on a multiple scale approach that is similar in spirit to turbulence transport theories~\citep{marsch_1989,zhou_1990, zank_1996,Breech_2005,Breech,zank_2017}. However, while the latter are designed to describe the radial evolution of a limited number of moments of the fluctuations (e.g. energy and cross helicity) and require closure assumptions, the resulting nonlinear equations for the fluctuations in the AEB can be solved via direct numerical integration in the frame co-moving with the solar wind.}  {In this sense our model differs from the one adopted by~\cite{nariyuki_2015}, who uses an accelerating expanding box for studying nonlinear Alfv\'en wave evolution in the Derivative Nonlinear Schr\"odinger approximation: in analogy with the traditional EBM, \cite{nariyuki_2015} derives a set of MHD equations that includes also the effects of the radial acceleration of the wind. However, this description does not include the expansion effects that arise from the gradients of background fields other than the wind speed --- in particular gradients of the Alfv\'en velocity which are non negligible in the acceleration region; second, contrary to the present model,  \cite{nariyuki_2015} considers a uniformly accelerating wind.} 

The inclusion of the acceleration region into the model, and in particular of the sub-Alfv\'enic corona, is important because in that region the strongest wave reflections occur and fluctuations' amplitudes are expected to reach their maximum -- both  central elements for the development of a turbulent cascade~\citep{velli_PRL_1989, matthaeus_ApJ_1999, verdini_ApJ_2009, perez_ApJ_2013}. Simulations of evolving fluctuations from the acceleration region out to the inner heliosphere is of further interest in view of the upcoming missions Solar Probe Plus~\citep{fox} and Solar Orbiter~\citep{muller}.  In particular, Solar Probe Plus will arrive closer to the Sun than any previous spacecraft by  crossing the  Alfv\'en critical point, placed somewhere between $10-20\,R_s$~\citep{deforest}, providing for the first time in-situ evidence of the state of the turbulence in the sub-Alfv\'enic corona. 

The paper is organized as follows: in Section~\ref{comoving} we introduce the basic idea of the  (accelerating) expanding box; in Section~\ref{average_wind} we write the equations describing the average solar wind that represents our resulting solar wind model; in Section~\ref{perturbations_equations} we derive the AEB governing equations for fluctuations and in Section~\ref{summary_calculations} we write down the governing equations of the AEB in terms of generalized Els\"asser variables used for numerical integration; in Section~\ref{properties} we discuss the range of validity of the AEB and a numerical test of Alfv\'en wave propagation, while the final discussion and summary is deferred to Section~\ref{discussion}. 

Here we lay emphasis mostly on propagation of MHD waves to validate the model. Nevertheless, the latter applies to a wider class of problems involving some localized structure, such as corotating stream interaction regions at sector boundaries of fast-slow wind~\citep{grappin_JGR_1996} and current sheets in the streamer belt~\citep{rappazzo_2005}. Finally, the model also provides a general approach to describe the dynamics of a given parent system of equations in an expanding, accelerating medium.

\section{The Accelerating Expanding Box}
\label{comoving}

Start from MHD with a polytropic energy equation of index $\gamma$, within the absolute frame of reference $\{x,y,z\}$ with origin in the Sun defined by
\begin{equation}
\begin{split}
&x=r\sin\theta\cos\varphi, \\ 
& y=r\sin\theta\sin\varphi,\\
& z=r\cos\theta,
\label{sph_cart}
\end{split}
\end{equation}
$r$, $\theta$, and $\varphi$ being the radial distance from the Sun, the latitude, and the longitude of a given point of the plasma, respectively. In this frame, the  set of equations for velocity, magnetic field, pressure and density is
\begin{subequations}
\begin{equation}
\begin{split}
\frac{\p {\bf V}}{\p t}+{\bf V\cdot\bn}{\bf V}=&-\frac{1}{\rho}\bn \left(  p +\frac{1}{8\pi}  B^2 \right)\\
&-\frac{G M_\odot }{r^2}{\bf \hat r}+\frac{1}{4\pi\rho}{\bf B\cdot\bn  B},
\end{split}
\label{eqSW}
\end{equation}
\begin{equation}
\frac{\p {\bf  B}}{\p t}+{\bf V\cdot\bn}{\bf B}=-{\bf  B(\bn\cdot V)}+{\bf  B\cdot\bn V},
\end{equation}
\begin{equation}
\frac{\p { p}}{\p t}+{\bf V\cdot\bn}{  p}=-\gamma  p(\bn\cdot {\bf V)},
\end{equation}
\begin{equation}
\frac{\p { \rho}}{\p t}+\bn\cdot(\rho {\bf V})=0.
\end{equation}
\label{eq_tot}
\end{subequations}

From eqs.~(\ref{eq_tot}) it is possible to obtain, through Reynolds average, two coupled set of equations describing respectively fluctuations and a spherically symmetric, statistically steady background  (see e.g. \cite{whang, zhou_1990}). Within this standard framework, we employ  a two-scale analysis by assuming that the average fields, labelled with a ``zero",  are functions of a slow radial variable $R=(\ell/R_0)\,r$, with $R_0$ a reference heliocentric distance and $\ell$ a dynamical fast scale length such that $\ell/R_0\sim\varepsilon\ll1$. The parameter $\varepsilon$ will be {\it de facto} our power expansion parameter. Fluctuations, on the other hand, are in general functions of $R$, via couplings with the background, of a fast spatial variable ${\bf r}$ and  of time $t$. Fields may therefore be separated into an average part, {steady in time and slowly varying in space}, and a fluctuating part as follows,

\begin{equation}
{\bf V}(R,{\bf r},t)={\bf U}(R)+{\bf u}(R,{\bf r},t),
\end{equation}
\begin{equation}
 {\bf B}(R,{\bf r},t)={\bf B_0}(R)+{\bf  b}(R,{\bf r},t)
\end{equation}
\begin{equation}
p(R,{\bf r},t) =p_0+ {\delta p}(R,{\bf r},t),
\end{equation}
\begin{equation}
 \rho(R,{\bf r},t)= \rho_0(R)+\delta\rho(R,{\bf r},t).
\end{equation}

Note that introduction of $R$ is a useful artifice, but $R$ and $r$ are not two independent variables (see e.g.~\cite{bender}), so that the radial derivative can be inferred from the usual chain rule for differentiation, 

\begin{equation}
\frac{d}{d r}=\frac{\partial}{\partial r}+\frac{\ell}{R_0}\frac{\partial}{\partial R}\equiv \frac{\partial}{\partial r}+\varepsilon\frac{\partial}{\partial R}.
\label{largescale}
\end{equation}

Hereafter it will be implied that gradients along the radial direction are given by both the small and the large scale variations as expressed by eq.~(\ref{largescale}).

{Reynolds average over}  eqs.~(\ref{eq_tot}) leads to the following solar wind equation,

\begin{equation}
\begin{split}
\rho_0&{\bf U\cdot\boldsymbol\nabla U}=- \boldsymbol\nabla p_0-\rho_0\frac{M_\odot G}{r^2}{\bf \hat r}+\frac{1}{4\pi}\left( \boldsymbol\nabla\times {\bf B}_0  \right)\times{\bf B}_0 \\
&+\boldsymbol\nabla\colon\left(\frac{1}{4\pi}<{\bf b b}>-<{\rho\bf uu}>-\frac{1}{8\pi}<b^2> \mathcal{I}\right)\\
&-<\delta\rho\frac{\partial{\bf u}}{\partial t}>-<\delta\rho({\bf u\cdot\bn U}+{\bf U\cdot\bn u})>,
\label{eq_U}
\end{split}
\end{equation}

where  brackets  denote time-average,  $\mathcal I$ is the unit dyadic, $M_\odot$  the  solar mass  and $G$ the gravitational constant. Equation~(\ref{eq_U}) is complemented by the equations for the background magnetic field, pressure and density,

\begin{subequations}
\begin{equation}
\begin{split}
&{\bf U\cdot\bn}{\bf B_0} =-{\bf  B_0(\bn\cdot U)}+{\bf  B_0\cdot\bn U}\\
&-\bn\colon\left({\bf<ub>-<bu}>\right),
\label{eq_b0}
\end{split}
\end{equation}
\begin{equation}
{\bf U\cdot\boldsymbol\nabla}p_0=-\gamma p_0{\bf(\boldsymbol\nabla\cdot U)}-\gamma< \bn\cdot(\delta p {\bf u} )>,
\end{equation}
\begin{equation}
{\bf U\cdot\boldsymbol\nabla} \rho_0=-\rho_0({\bf \boldsymbol\nabla\cdot U)}-<\bn\cdot(\delta\rho {\bf u})>,
\label{eq_p0}
\end{equation}
\label{back0}
\end{subequations}

and by the governing equations for the fluctuations, obtained by subtracting eqs.~(\ref{eq_U})--(\ref{back0}) from eqs.~(\ref{eq_tot}).

The equations for the fluctuations  are best written in terms of the normalized magnetic field and pressure
 
\begin{equation}
{\bf \tilde b}=\frac{\bf b}{\sqrt{4\pi\rho_0}},\qquad   \tilde p=\frac{p}{c_s \rho_0},
\end{equation}

where $c_s=\sqrt{\gamma p_0/\rho_0}$ is the sound speed. We write them below by denoting with a ``prime" the large scale radial derivative $\partial/\partial R$, and where $\delta=\delta\rho/\rho_0$ and  ${\bf V_a}={\bf B_0}/\sqrt{4\pi\rho_0}$ is the Alfv\'en velocity, 
 
 \begin{subequations}
\begin{equation}
\begin{split}
\frac{\p {\bf u}}{\p t}+&{\bf U\cdot\bn}{\bf u}+{\bf u\cdot\bn}{\bf U}+{\bf u\cdot\bn}{\bf u}-{\bf V_a\cdot\bn \tilde b}\\
&-{\bf\tilde b\cdot\bn  V_a}-{\bf\tilde b\cdot\bn \tilde b}=\frac{1}{2}{\bf \tilde b}(V_{a,r}+\tilde b_r)(\ln \rho_0)^\prime \\
&+\frac{1}{2}{\bf V_a}\tilde b_r(\ln \rho_0)^\prime-\bn \tilde P_T\\
&-\frac{\delta}{1+\delta}\left[ {\bf\tilde b\cdot\bn \tilde  b}+{\bf\tilde b\cdot\bn V_a}+{\bf V_a\cdot\bn \tilde  b}+{\bf V_a\cdot\bn V_a}\right]\\
&-\frac{\delta}{1+\delta}\left[\frac{1}{2}({\bf \tilde b+V_a})(V_{a,r}+\tilde b_r)(\ln \rho_0)^\prime\right]-\langle\dots\rangle,
\end{split}
\label{mot}
\end{equation}
\begin{equation}
\begin{split}
\frac{\p {\bf \tilde b}}{\p t}+&{\bf U\cdot\bn}{\bf\tilde b}+{\bf u\cdot\bn}{\bf\tilde b}\\
&+{\bf u\cdot\bn}{\bf V_a}-{\bf \tilde b\cdot\bn U}-{\bf \tilde b\cdot\bn u}-{\bf  V_a\cdot\bn u}=\\
&-\frac{1}{2}{\bf \tilde b}(U+u_r) (\ln \rho_0)^\prime -\frac{1}{2}{\bf V_a}u_r (\ln \rho_0)^\prime \\
&-{\bf \tilde b(\bn\cdot U)}-{\bf \tilde b(\bn\cdot u)}-{\bf V_a (\bn\cdot u)}-\langle\dots\rangle,
\label{far}
\end{split}
\end{equation}
\begin{equation}
\begin{split}
\frac{\p {\delta\tilde{ p}}}{\p t}&+{\bf U\cdot\bn}{\delta \tilde{ p}}+{\bf u\cdot\bn}{\delta \tilde{ p}}+u_r\frac{1}{\gamma}c_s^\prime \\
&+\frac{c_s}{\gamma}u_r (\ln\rho_0 c_s)^\prime +\delta\tilde{ p}(U+ u_r) (\ln\rho_0 c_s)^\prime=\\
&-(c_s+\gamma \delta\tilde{ p})(\bn\cdot {\bf u)} -\gamma\delta\tilde{ p} (\bn\cdot {\bf U)}-\langle\dots\rangle,
\end{split}
\end{equation}
\begin{equation}
\begin{split}
&\frac{\p }{\p t}\left(\frac{\delta\rho}{\rho_0}\right)+{\bf U\cdot\bn}\left(\frac{\delta\rho}{\rho_0}\right)+{\bf u\cdot\bn}\left(\frac{\delta\rho}{\rho_0}\right)\\
&+U\frac{\delta\rho}{\rho_0}(\ln \rho_0)^\prime+u_r \left(\frac{\delta\rho}{\rho_0}+1\right)(\ln\rho_0)^\prime =- \left(\frac{\delta\rho}{\rho_0}\right)({ \bf\bn\cdot U)} \\
&- \left(\frac{\delta\rho}{\rho_0}+1\right)({ \bf\bn\cdot u)}-\langle\dots\rangle.
\end{split}
\end{equation}
\label{eqflu}
\end{subequations}

Again,  brackets $\langle\dots\rangle$ stand for the nonlinear feedback on the average fields that must be subtracted, whereas the gradient of the  total pressure $\tilde P_T$ in eq.~(\ref{mot}) is given by

\begin{equation}
\begin{split}
 \bn \tilde P_T= \frac{1}{\rho}\bn & \left[ \delta \tilde p(\rho_0c_s)+\frac{1}{2}\rho_0(\tilde b^2+2{\bf \tilde b\cdot V_a})\right]\\
 &-\frac{1}{\rho_0}\frac{\delta}{1+\delta}\bn \left(p_0+\frac{1}{2}V_a^2\right).
 \end{split}
\end{equation}

\begin{figure*}[t]
\begin{center}
\includegraphics[width=0.6\textwidth]{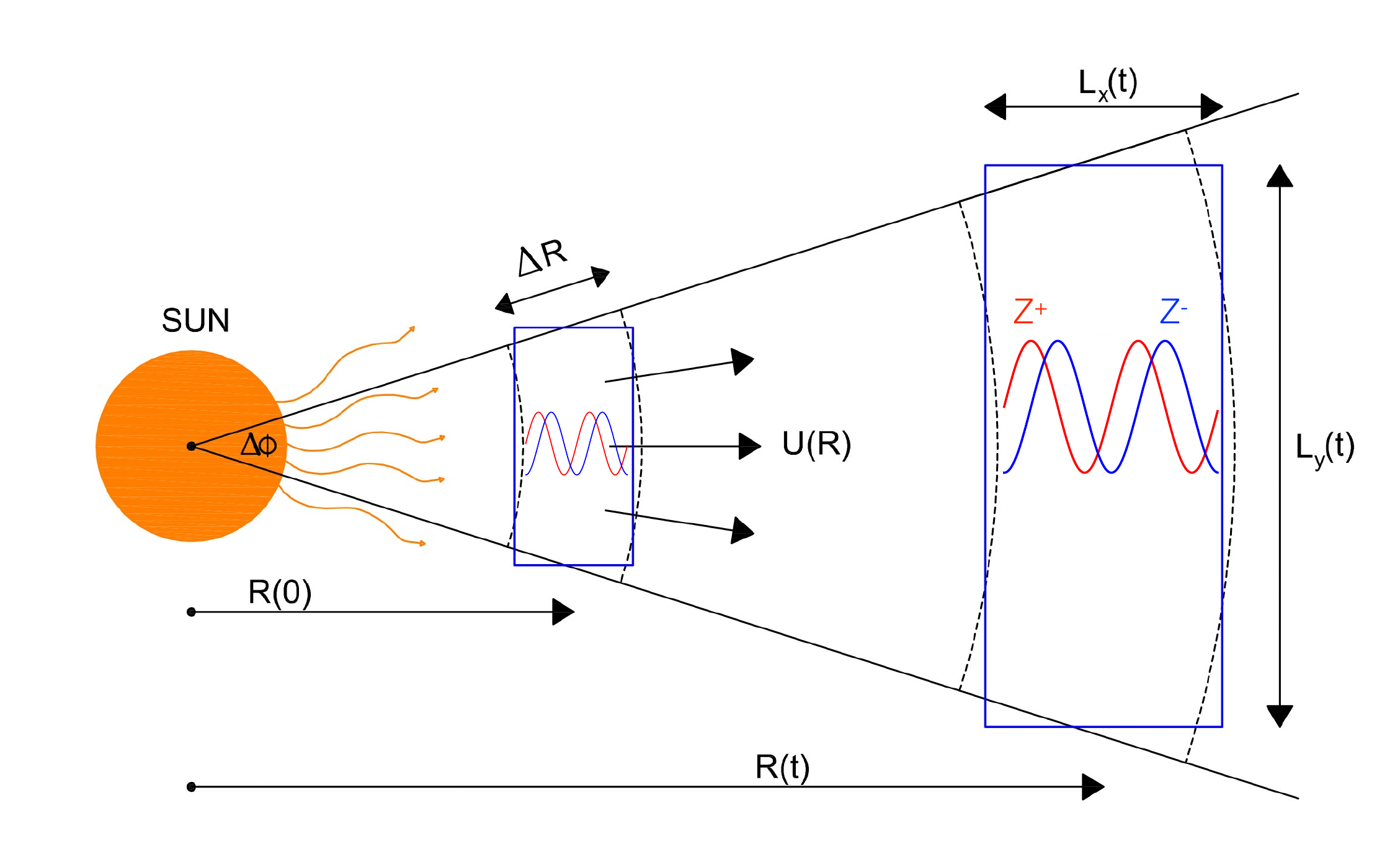}
\caption{Sketch of the accelerating expanding box. The cartesian co-moving frame is represented by the blue boxes which stretch as the average heliocentric distance $R(t)$ increases with time following the radial expansion.}
\label{aebimage}
\end{center}
\end{figure*}

The aim of the AEB is  to describe how inhomogeneity and expansion affect the radial evolution of fluctuations given in eqs.~(\ref{eqflu}). For this reason, the AEB does not take into account the nonlinear feedback of fluctuations on the average quantities in eqs.~(\ref{back0}), and the  profile of ${\bf U}$ is given {\it a priori} in this model. The main  idea is  to change to a new frame of reference, the co-moving frame, that allows to follow in a semi-lagrangian fashion the evolution of a small angular sector of plasma (the simulation box) which is in accelerating, spherical expansion because of the drag of the average outflow ${\bf U}=U(R){\bf\hat{r}}$. A schematic representation of the AEB  is shown in Fig.~\ref{aebimage}. In this model the expansion is taken into account by introducing an explicit dependence on the variable $R(t)$, which is defined here as the average heliocentric distance of the plasma sector, taken along the $x$ axis.  The average distance $R(t)$ is parameterized in time and can be considered the lagrangian variable of the background, obeying to the following equation,
\begin{equation}\frac{dR}{dt}=U(R).
\end{equation} 

At any given time, a point of the plasma parcel has coordinates defined by $R(t)$ and by the local position vector ${\delta\bf{ r}}$,

\begin{equation}
{\delta\bf {r}}=(\delta x,\,y,\,z),\quad \delta x=x-R(t).
\end{equation}

As can be seen from eqs.~(\ref{back0}) and (\ref{eqflu}), effects of the solar wind expansion on the temporal evolution of scalar and vectorial fields enter explicitly through the convective derivative associated with the flow  ${\bf U\cdot\boldsymbol\nabla}$,  its divergence ${\bf\boldsymbol\nabla\cdot U}$, and, for a generic vector ${\bf A}$,  the material derivative ${\bf A\cdot\boldsymbol\nabla U}$.  If the plasma volume  is small, i.e., if the  length of the box is of order $\delta x/R\sim \epsilon\ll1$ and the transverse dimensions are of  order $ y/R,\, z/R=\alpha\ll1$, at any given time one can expand the spherical coordinates  around $R$ to first order in the small parameters  $\alpha$ and~$\epsilon$, so as to obtain an explicit dependence on $R$. This procedure allows us to include the expansion   effects on the local dynamics.   The difference with the  EBM consists  in retaining the first correction due to the acceleration of $U(r)$,  that gives an additional contribution to the expansion rates, as we will see. 

In practice, if the radial inhomogeneity of the solar wind speed $U(r)$ is retained up to  first order derivative, $U(r)$ can be written locally  as
\begin{equation}
{U}(r)=U(R)+U^\prime \delta x+\mathcal{O}(\epsilon^2,\alpha^2),
\label{diff_U}
\end{equation}

where $U^\prime=({dU}/{dr})|_R$  has at most the order of magnitude $U^\prime\sim U/R$.  The components of ${\bf U}$ are given by
\begin{equation}
{\bf U}(r)=U(r)\left[\frac{\delta x +R}{r}\right]{\bf\hat{x}}+U(r)\frac{ y}{r}{\bf\hat{y}} +U(r)\frac{z}{r}{\bf\hat{z}},
\end{equation}
which, together with eq.~(\ref{diff_U}), yields to first order in $\alpha,\epsilon$:

\begin{equation}
{\bf U}(R)=\left[U(R)+U^\prime \delta x\right]{\bf\hat{x}}+U(R)\frac{ y}{R}{\bf\hat{y}} +U(R)\frac{z}{R}{\bf\hat{z}}.
\label{U_noncom}
\end{equation}

According to the spherical expansion expressed by eq.~(\ref{U_noncom}), the plasma parcel undergoes a stretching in the transverse directions $y$ and $z$ at a rate $U/R$  because of the advection of the solar wind velocity ${\bf U}$. On the other hand, while considering a finite radial acceleration within the box,  there is also an additional stretching along the parallel (radial) direction. Since we are assuming that the  length of the box $L_x$ is small, $L_x/R\ll 1$, the latter increases at a rate $dL_x/dt = L_xU^\prime$. 

The co-moving frame is thus defined by the following change of variables:
\begin{equation}
\begin{split}
&\bar{x}=[x-R(\bar t)]\left({L_{0x}}/{L_x(t)}\right),\\
& \bar{y}=y\left({R_0}/{R(t)}\right),\\
& \bar{z}=z\left({R_0}/{R(t)}\right),\\
& \bar{t}=t,
\end{split}
\label{com}
\end{equation}
\begin{equation}
\frac{dR}{d\bar t}=U,\quad \frac{dL_x}{d\bar t}=L_xU^\prime.
\end{equation}

In the equations above,  $R_0$ and $L_{0x}$ are the initial average heliocentric distance and parallel length of the box, respectively. It is easily verified that fast spatial derivatives transform as 

\begin{equation} 
\boldsymbol\nabla=\frac{L_{0x}}{L_x}\frac{\partial}{\partial{\bar x}}{\bf\hat x}+ \frac{R_0}{R(t)}\frac{\partial}{\partial{\bar y}}{\bf\hat y}+ \frac{R_0}{R(t)}\frac{\partial}{\partial{\bar z}}{\bf\hat z}\equiv\bar{\boldsymbol\nabla},
\label{deriv_com}
\end{equation}

whereas the slow spatial variation along the radial direction, that we have introduced in eq.~(\ref{largescale}), becomes a slow time variation as follows 

\begin{equation} 
\varepsilon\frac{\partial}{\partial{R}}=\varepsilon\frac{1}{U}\frac{\partial}{\partial{\tau}}.
\label{deriv_t_com}
\end{equation}

We recall however that in reality fields are functions of $t$ alone, and $\tau$ has to be ultimately included within the variable~$t$.

The solar wind velocity can be written, in the new frame,

\begin{equation} 
\begin{split} 
{\bf \bar U}\left[U(R)+U^\prime \left(\frac{L_x}{L_{0x}}\right)\bar x\right]{\bf\hat{x}}+U(R)\frac{\bar y}{R_0}{\bf\hat{y}} +U(R)\frac{\bar z}{R_0}{\bf\hat{z}},
\end{split} 
\end{equation}
and consequently the  time derivative is given by

\begin{equation} 
\frac{\partial}{\partial{t}}=\frac{\partial}{\partial{\bar t}}-{\bf \bar U\cdot \boldsymbol{\bar\nabla}}.
\label{deriv_t_com}
\end{equation}

The time-dependent rescaling of coordinates defined in eqs.~(\ref{com}) simplifies the convective derivative associated with ${\bf U}$, while the slow radial gradient implicitly becomes a slow time variation, $U(\p /\p R)=\partial/\partial\tau$. In this way, any dependence on the local coordinates due to the advection is eliminated, and we can assume a background  locally homogeneous, yet slowly varying in time, which allows for the use of periodic boundary conditions. Finally, the divergence of ${\bf U}$ and the material derivative of a generic vector ${\bf A}$ can be written in the co-moving frame, according to our ordering assumptions, in the following way:

\begin{equation}
{\bf{\boldsymbol\nabla}\cdot  U}=\frac{2U(R)}{R(t)}+U^\prime(R),
\label{divU}
\end{equation}
\begin{equation}
{\bf A\cdot{\boldsymbol\nabla} U}(R)=U^\prime(R) \bar A_{x}{\bf\hat{{x}}}+\frac{U(R)}{R(t)} \bar A_{y}{\bf\hat{{y}}} +\frac{U(R)}{R(t)} \bar A_{z}{\bf\hat{{z}}}. 
\label{materialU}
\end{equation}

In summary, the expansion effects in the co-moving frame $\{\bar x,\,\bar y,\,\bar z\}$ are represented  by: $(i)$ the rescaling of spatial derivatives, described by eqs.~(\ref{com});  $(ii)$ the divergence of the solar wind velocity, eq.~(\ref{divU}); $(iii)$  the material derivative of ${\bf U}$ with respect to a given vectorial field, eq.~(\ref{materialU}). 

The  quantities expressed by eqs.~(\ref{divU})--(\ref{materialU})  are the  expansion terms which appear as effective drag terms in the reduced set of equations to be  discussed in the following sections. These terms introduce an expansion time scale $\tau_e=R_0/U_\infty$ which is assumed large with respect to the local dynamical time scale, where $U_\infty$ is a reference value for $U(R)$, e.g. its  asymptotic  value. The new expansion terms include effects of  the wind acceleration and are an extension of those already present in the EBM.

However,  within the acceleration region, where the solar wind crosses the critical sound and Alfv\'en points,  the coupling of fluctuations with the  inhomogeneity of background quantities other than ${\bf U}$, e.g. background density, sound and Alfv\'en speed, may not be neglected. Indeed,  this coupling  introduces corrections of  order of $c_s/U$ and $V_a/U$ with respect to the expansion terms.  For instance, take  the induction equation (\ref{far}):  the term which couples fluctuations and  gradient of the background magnetic field is  $-{\bf u\cdot\boldsymbol\nabla V_a}\sim u(V_a/R)$. It can be seen easily that the latter is of order  $V_a/U$ with respect to the expansion term ${\bf \tilde b}(\boldsymbol\nabla\cdot {\bf U})\sim \tilde b(U/R)$,  assuming an Alfv\'enic correlation of the type $u\sim\tilde b$. As a consequence, if one wants to  describe correctly plasma dynamics in the acceleration region near the critical Alfv\'en point where $V_a=U$, this term must be retained.  Similar considerations hold for the pressure, continuity, and momentum equations where the additional coupling of fluctuations with the large scale gradients of the sound speed  introduces terms $\sim c_s/U$. As we will discuss, all these large scale gradient terms that we introduce in this model are the first necessary corrections to be kept in order to conserve the wave action flux of small amplitude waves. In particular, we will introduce the simplest form of such large scale gradients necessary to conserve the generalized wave action flux of Alfv\'en waves, and we will consider the case $c_s/U\ll 1$.   

\section{The average solar wind}
\label{average_wind}
We prescribe a  profile of the solar wind speed $U(R)$, so that the background mass density, (non normalized) pressure and magnetic field evolve self-consistently according to eqs.~(\ref{back0}) together with eqs.~(\ref{divU})--(\ref{materialU}). Taking $R_0$ as a reference heliocentric distance and indicating with a  ``star" quantities  at $R_0$, we obtain the following radial evolution for density, pressure and magnetic field, that we label with an over-bar:

\begin{widetext}
\begin{subequations}
\begin{equation}
\frac{1}{\bar\rho_0}\frac{d\bar \rho_0}{dR}=-\left( \frac{2}{R}+\frac{U^\prime}{U}\right)\quad\Rightarrow\quad\bar \rho_0=\rho_*\left(\frac{R_0}{R}\right)^{2}\frac{U_*}{U},
\end{equation}
\begin{equation}
\frac{1}{\bar p_0}\frac{d\bar p_0}{dR}=-\gamma\left( \frac{2}{R}+\frac{U^\prime}{U}\right)\quad\Rightarrow\quad\bar p_0=p_*\left(\frac{R_0}{R}\right)^{2\gamma}\frac{U_*}{U},
\end{equation}
\begin{equation}
\frac{1}{\bar B_{0,x}}\frac{d\bar B_{0,x}}{dR}=- \frac{2}{R}\quad\Rightarrow\quad \bar B_{0, x}=B_{* x}\left(\frac{R_0}{R}\right)^{2},\,\bar V_{a,x}=\frac{\bar B_{0,x}}{\sqrt{\bar\rho_0}},
\end{equation}
\begin{equation}
\frac{1}{\bar B_{0,y}}\frac{d\bar B_{0,y}}{dR}=-\left( \frac{1}{R}+\frac{U^\prime}{U}\right)\Rightarrow\quad \bar B_{0, y}=B_{* y}\left(\frac{R_0}{R}\right)\frac{U_*}{U},\,\bar V_{a,y}=\frac{\bar B_{0,y}}{\sqrt{\bar\rho_0}}.
\end{equation}
\label{back1}
\end{subequations}
\end{widetext}

By introducing the solar wind speed and  $R(\bar t)$, the former slow radial dependence in $R/R_0$ in the fixed frame of reference becomes an explicit slow time dependence in $\bar t/\tau_e$, $\tau_e=R_0/U_\infty$ being the expansion time scale. For example,  consider the asymptotic regime of expansion,  where $U=U_\infty$ and

\begin{equation}
R(\bar t)=R_0+U_\infty\, \bar t=R_0(1+\bar t/\tau_e).
\end{equation}
In this case,  density and  pressure   decrease respectively as 
\begin{equation}
\bar \rho_0\propto (R_0/R)^2=\frac{1}{(1+\bar t/\tau_e)^2},
\end{equation}
\begin{equation}
\bar  p_0\propto(R_0/R)^{2\gamma}=\frac{1}{(1+\bar t/\tau_e)^{2\gamma}},
\end{equation}

according to the conservation of mass flux and adiabatic cooling, as expected in our model for the solar wind. At small heliocentric distances in the acceleration region the expansion rates are larger, and density and pressure decrease faster. 

Within the same regime we get also the following expressions for the magnetic field:

\begin{equation}
 \bar B_{0,x}\propto (R_0/R)^2,\quad  \bar B_{0,y}\propto (R_0/R).
\label{B_ebm}
\end{equation}

The scaling of the angle of the Parker spiral with heliocentric  distance, which is $\propto R/R_0\propto \bar t/\tau_e$, is thus recovered. The set given by eqs.~(\ref{back1}) for given $U(R)$ will represent our  average solar wind. Note that at our order of  expansion of $U(r)$, the magnetic field component $\bar B_{0,x}$ and $\bar B_{0,y}$ respectively mimic  the radial and the azimuthal component of the average magnetic field. 

\section{Derivation of the AEB governing equations}

\label{perturbations_equations}

We want to approximate  eqs.~(\ref{eqflu}) in the co-moving frame in order to describe some fundamental properties of fluctuations -- in particular  of   Alfv\'en waves -- related to  solar wind inhomogeneity and expansion, while keeping a locally homogeneous and plane geometry. 

In order to obtain the evolution equations in the AEB from eqs.~(\ref{eqflu}) it is therefore necessary not only to introduce explicitly the expansion terms (${\bf A\cdot \bn U}$ and $\bn\cdot {\bf U}$), but also to write, as a function of  the known quantities $U$, $U^\prime$ and $R$, all spatial derivatives of the other background fields (${\bf V_a}$, $\rho_0$, $c_s$) before changing to the co-moving frame. If not, such important terms which couple fluctuations with the gradients of the average solar wind would be lost, as the background is, by construction, locally homogeneous in the (accelerating) expanding box approximation. We will however   retain only the first large scale gradient corrections necessary to  describe correctly Alfv\'en waves in the so-called non-WKB approximation, in which the generalized wave action flux, involving both ``outwards" and ``inwards" waves, must be conserved. To this end, we rely on the following approximations: $(i)$  we assume a radial background magnetic field, since, as explained, the large scale gradient corrections are non-negligible only close the Alfv\'enic point, where the spiral angle is very small, but equations can be generalized to include terms $\sim V_{a,\varphi}/U$; $(ii)$ we consider the supersonic region and neglect large scale gradient terms of order $c_s/U$ involving compressible fluctuations, as well as those large scale gradients in the momentum equation coupling with density perturbations -- inclusion of these effects on compressible fluctuations may be deferred to a later work; $(iii)$ we retain the remaining large scale gradients up to first order in the fluctuations' amplitude. We therefore neglect those  terms in the $\bn \tilde  P_T$  since the latter is essentially nonlinear for  Alfv\'en waves. 

Within these assumptions, we obtain the following equations valid in the Sun-centered fixed frame of reference -- notice that the large scale advection at the solar wind and Alfv\'en speed, of the form $U(\p/\p R)$ and $V_a(\p/\p R)$,  are implicitly included in the $\bn$ operator:

\begin{widetext}
\begin{subequations}
\begin{equation}
\begin{split}
\frac{\p {\bf u}}{\p t}+&{\bf U\cdot\bn}{\bf u} -{\bf V_a\cdot\bn \tilde b}+{\bf u\cdot\bn}{\bf u}-{\bf\tilde b\cdot\bn \tilde b}=\frac{1}{2}\left({\bf \tilde b}V_{a,r}+{\bf V_a}\tilde b_r\right)(\ln \rho_0)^\prime\\
&-{\bf u}\cdot\bn{\bf U}+{\bf\tilde b\cdot\bn V_a} -\bn \tilde P_T-\frac{\delta}{1+\delta}\left( {\bf\tilde b\cdot\bn \tilde  b}+{\bf V_a\cdot\bn \tilde  b}\right),
\end{split}
\end{equation}
\begin{equation}
\begin{split}
\frac{\p {\bf \tilde b}}{\p t}+&{\bf U\cdot\bn}{\bf \tilde b}-{\bf  V_a\cdot\bn u}+{\bf u\cdot\bn}{\bf\tilde b}-{\bf \tilde b\cdot\bn u}=\\
&-{\bf u\cdot\bn V_a}+{\bf \tilde b\cdot\bn U}-\frac{1}{2}\left({\bf \tilde b}U+{\bf V_a}u_r\right) (\ln \rho_0)^\prime-{\bf \tilde b(\bn\cdot U)}-{\bf \tilde b(\bn\cdot u)}-{\bf V_a (\bn\cdot u)},
\end{split}
\label{ind}
\end{equation}
\begin{equation}
\frac{\p {\delta\tilde p}}{\p t}+{\bf U\cdot\bn}{\delta \tilde p}+{\bf u\cdot\bn}{\delta \tilde p}=-\left(c_s+\gamma \delta\tilde p\right)(\bn\cdot {\bf u)}-\left(\frac{\gamma-1}{2}\right)\delta\tilde p (\bn\cdot {\bf U)},
\end{equation}
\begin{equation}
\frac{\p }{\p t}\left(\frac{\delta\rho}{\rho_0}\right)+{\bf U\cdot\bn}\left(\frac{\delta\rho}{\rho_0}\right)+{\bf u\cdot\bn}\left(\frac{\delta\rho}{\rho_0}\right) =- \left(\frac{\delta\rho}{\rho_0}+1\right)(\bn\cdot{ \bf u)},
\end{equation}
\begin{equation}
\begin{split}
 \bn \tilde P_T= \frac{1}{\rho}\bn  \left[ \delta\tilde p(\rho_0c_s)+\frac{1}{2}\rho_0(\tilde b^2+2{\bf \tilde b\cdot V_a})\right].
\end{split}
\end{equation}
\label{sph}
\label{aeb_one}
\end{subequations}
\end{widetext}

The gradients of  the Alfv\'en velocity can now be taken into account explicitly by expanding its components to first order in $\epsilon$ and $\alpha$, in the same way we expanded the solar wind velocity in eq.~(\ref{U_noncom}). The material derivative of ${\bf V_a}$, for a generic vector ${\bf A}$, therefore reads 

\begin{equation}
\begin{split}
&{\bf A\cdot\bn V_a}=  -A_x\left(\frac{\bar V_{a,x}}{R}-\frac{1}{2}\bar V_{a,x}\frac{U^\prime}{U}\right){\bf \hat x}\\
&+A_y\frac{\bar V_{a,x}}{R}{\bf \hat y}+A_z\frac{\bar V_{a,x}}{R}{\bf \hat z},
\label{grad}
\end{split}
\end{equation}

with

\begin{equation}
\bar V_{a,x}=V_{a*,x}\frac{R_0}{R}\sqrt{\frac{U}{U_*}}.
\end{equation}

We now change to the co-moving frame by using the coordinate transformation (\ref{com}) and (\ref{deriv_t_com}), eqs.~(\ref{divU})--(\ref{materialU}) and eq.~(\ref{grad}), and retain the simplest Alfv\'enic couplings introduced by inhomogeneity that reproduce the known properties of Alfv\'en waves, in a way which allows to conserve the condition $\bn\cdot {\bf b}=0$. We therefore impose that the coupling introduced by the Alfv\'en speed inhomogeneity be between the magnetic field and the incompressible part of the velocity. A more detailed discussion on the problem of the conservation of the zero-divergence of the magnetic field, arising from our simplification of the term ${\bf \bn \times (u\times B_0)}$, can be found in Appendix~\ref{LSC_induction}.

Since the background fields in the AEB are homogeneous in space, we can recast the system of equations in a more compact form. We obtain in this way the AEB  model,  complemented by eqs.~(\ref{back1}): 

\begin{widetext}
\begin{subequations}
\begin{equation}
\begin{split}
\frac{\p {\bf u}}{\p t}&- \frac{V_{a,x}}{U}\frac{\p {\bf \tilde b}}{\p t} +{\bf u\cdot\bn}{\bf u}=-\frac{1}{\rho}\bn \left[ \delta\tilde{ p} (\rho_0 c_s)+\frac{\rho_0}{2}(\tilde b^2+2{\bf \tilde b\cdot V_a}) \right]+\frac{\rho_0}{\rho}{\bf( \tilde b+V_a)\cdot\bn \tilde b}\\
&-\hat P_u\cdot{\bf u}+{\bf\tilde b_\bot} \frac{V_{a,x}}{R}-\frac{1}{2}{\bf \tilde b}V_{a,x}\left(\frac{2}{R}+\frac{U^\prime}{U} \right),
\end{split}
\end{equation}
\begin{equation}
\begin{split}
\frac{\p {\bf \tilde b}}{\p t}&-\frac{V_{a,x}}{U}\frac{\p {\bf u^{\text{inc.}}}}{\p t}=\bn\times[{\bf u\times(\tilde b+V_a)]}-\hat P_b\cdot {\bf \tilde b}-{\bf u_\bot^{\text{inc.}}}\frac{V_{a,x}}{R}-{\bf \hat x}\,u_x^{\text{inc.}}V_{a,x}\frac{U^\prime}{U}+\frac{1}{2}{\bf \tilde b} U\left(\frac{2}{R}+\frac{U^\prime}{U} \right),
\end{split}
\label{ind_inc}
\end{equation}
\begin{equation}
\frac{\p {\delta\tilde p}}{\p t}+{\bf u\cdot\bn}{\delta \tilde p}=-\left(c_s+\gamma \delta\tilde p\right)(\bn\cdot {\bf u)}-\left(\frac{\gamma-1}{2}\right)\delta\tilde p \left(\frac{2U}{R}+U^\prime \right),
\end{equation}
\begin{equation}
\frac{\p }{\p t}\left(\frac{\delta\rho}{\rho_0}\right)+{\bf u\cdot\bn}\left(\frac{\delta\rho}{\rho_0}\right) =- \left(\frac{\delta\rho}{\rho_0}+1\right)(\bn\cdot{ \bf u)}.
\end{equation}
\label{aeb1}
\end{subequations}
\end{widetext}

For the sake of notation  we have dropped the over-bar, and the projectors $\hat P_u$, $\hat P_b$ and the gradient are defined as follows:

\begin{equation}
\hat P_u=\left(U^\prime,\,\frac{U}{R},\,\frac{U}{R}\right),
\end{equation}
\begin{equation}
 \hat P_b=\left(\frac{2U}{R},\,\frac{U}{R}+U^\prime ,\,\frac{U}{R}+U^\prime \right),
\end{equation}
\begin{equation}
 \bn=\left(\frac{L_{0x}}{L_x}\frac{\p}{\p x},\,\frac{R_{0}}{R}\frac{\p}{\p y},\,\frac{R_{0}}{R}\frac{\p}{\p z}\right).
\end{equation}

In summary, equations~(\ref{aeb1}) contain the same expansion terms of the previous expanding box model~\citep{grappin_JGR_1996} extended to include effects of the wind acceleration. In addition, terms of order  $V_a/U$  with respect to the traditional expansion ones now arise in the induction and momentum equation, coupling the incompressible part of the velocity (${\bf u}^{\text {inc.}}$) and the magnetic field ${\bf \tilde b}$. Those terms are the large scale radial advection at the Alfv\'en speed, corresponding to the second term on the left-hand-side of either equations (this term originates from the implicit advection term $V_{a,r}(\p /\p R)$), and  the last two terms on the right-hand-side (essentially originating from the inhomogeneity of the Alfv\'en speed): these are the first necessary corrections to be kept in order to conserve Alfv\'en wave action flux and to describe correctly the radial evolution of the intrinsic frequency of waves at small distances from the Sun, as we will discuss in Section~\ref{normal}.  

However, as can be seen, the large scale advection at the Alfv\'en speed  along the radial direction couples the time derivatives of the incompressible part of ${\bf u}$ and ${\bf \tilde b}$ -- as it must be for Alfv\'en waves.  Although one might be tempted to neglect such a term, therefore greatly simplifying the temporal integration of the AEB equations,  Alfv\'en waves would be  described only qualitatively across the  Alfv\'en critical point. In particular, while the amplitude of outwards waves would still peak at the critical point according to the WKB approximation, it would display an exponential radial variation in the neighborhood of the critical point, rather than the algebraic one predicted by linear theory. The correction due to the large scale advection at the Alfv\'en speed was not taken into account in a previous version of the model~\citep{tenerani_JGR_2013}. In the next section we provide an equivalent form of the AEB which allows to tackle the problem of  the temporal integration across the  Alfv\'en critical point of the full equations given in~(\ref{aeb1}).  

\section{Model equations and numerical implementation}
\label{summary_calculations}

In order to integrate eqs.~(\ref{aeb1}) we need to decouple the induction and momentum equation by introducing generalized Elas\"asser variables involving the vorticity ${\bf \meta=\bn\times u}$ and the current density ${\bf \tilde j=\bn\times\tilde b}$. To this end, we make use of the Helmholtz decomposition ${\bf u=\boldsymbol\nabla\chi+\boldsymbol\nabla\times V}$, where the scalar and vector potentials $\chi$ and ${\bf V}$ are such that

\begin{equation}
{\bf \boldsymbol\nabla\times u}=-\nabla^2 {\bf V}\equiv \meta,\quad {\bn\cdot \bf  u}=\nabla^2\chi\equiv\xi.
\end{equation}

The sought set of equations can be found by taking the curl and the divergence of the momentum equation and the curl of the induction equation and, in analogy with the Els\"asser variables, by definining  ${ \meta^\mp}=\meta \mp{\bf \tilde j}$. Evidently, we  drop the large scale gradient terms in the compressible part of the velocity as this does not contribute to shear Alfv\'en modes. The AEB equations can therefore be rewritten for these new generalized Els\"asser variables, the divergence $\xi$, pressure and density as: 

\begin{widetext}
\begin{subequations}
\begin{equation}
\begin{split}
\left(1\pm\frac{V_{a,x}}{U}\right)&\frac{\p {\meta^\mp}}{\p t}+{\bf u\cdot\bn\meta^\mp}= \meta\cdot\bn{\bf u}\mp{\bf \tilde B}\cdot\bn \meta^\mp\pm\frac{V_{a,r}}{R} \meta^\pm-\frac{1}{2}\frac{ \meta^+- \meta^-}{2}\left(\frac{2}{R}+\frac{U^\prime}{U}\right)(V_{a,r}\pm U)\\
&{\bf \mp\{\tilde b,u\} \pm\{u, \tilde b\} }+\frac{\rho_0}{\rho} {\bf \{\tilde b,\tilde b\}}-\hat P_\eta  \frac{\meta^++ \meta^-}{2}\pm \hat P_J  \frac{\meta^+- \meta^-}{2}\\
&-\frac{\delta}{1+\delta}{\bf \tilde B\cdot\bn \tilde j}-\frac{\rho_0}{\rho^2}\bn\delta\rho\times({\bf\tilde B\cdot\bn\tilde b})\pm\bn\xi\times{\bf\tilde B}+\frac{1}{\rho^2}\bn\delta\rho\times\bn P_T-\xi\meta^\mp\mp\frac{V_a}{R}\eta_x^\mp{\bf \hat x}\\
&\frac{V_a}{U}\left[ -U^\prime\frac{\p \tilde b_z}{\p x}\mp\left( 2\frac{U}{R}\frac{\p u_x^{\text{inc}}}{\p z}-U^\prime\frac{\p u_z^{\text{inc}}}{\p x}-U^\prime\frac{\p u_x^{\text{inc}}}{\p z}\right) \right]{\bf \hat y}\pm\left( 2 \frac{U}{R}\frac{\p \tilde b_x}{\p z}-U^\prime\frac{\p \tilde b_z}{\p x}-U^\prime\frac{\p \tilde b_x}{\p z}\right){\bf \hat y}\\
&+\frac{V_a}{U}\left[ U^\prime\frac{\p \tilde b_y}{\p x}\mp\left( -2\frac{U}{R}\frac{\p u_x^{\text{inc}}}{\p y}+U^\prime\frac{\p u_y^{\text{inc}}}{\p x}+U^\prime\frac{\p u_x^{\text{inc}}}{\p y}\right)\right]{\bf \hat z}\pm\left( - 2\frac{U}{R}\frac{\p \tilde b_x}{\p y}+U^\prime\frac{\p \tilde b_y}{\p x}+U^\prime\frac{\p \tilde b_x}{\p y}\right){\bf \hat z},
\end{split}
\label{meta}
\end{equation}
\begin{equation}
\begin{split}
\frac{\p { \xi}}{\p t}+{\bf u\cdot\bn}\xi&+\sum_{i,j}(\p_i u_j)(\p_j u_i)=-\frac{1}{\rho}\nabla^2P_T+\frac{1}{\rho^2}\bn\delta\rho\cdot\bn P_T\\
&-\frac{\rho_0}{\rho^2}\bn\delta\rho\cdot[({\bf V_a+\tilde b})\cdot\bn {\bf\tilde b}]+\frac{\rho_0}{\rho}(\p_i \tilde b_j)(\p_j\tilde b_i)-2U^\prime\frac{\p u_x}{\p x}-2\frac{U}{R}\left( \frac{\p u_y}{\p y}+ \frac{\p u_z}{\p z}\right),
\end{split}
\end{equation}
\begin{equation}
\frac{\p {\delta\tilde p}}{\p t}+{\bf u\cdot\bn}{\delta \tilde p}=-\left(c_s+\gamma \delta\tilde p\right)\xi-\left(\frac{\gamma-1}{2}\right)\delta\tilde p \left(\frac{2U}{R}+U^\prime\right),
\end{equation}
\begin{equation}
\frac{\p }{\p t}\left(\frac{\delta\rho}{\rho_0}\right)+{\bf u\cdot\bn}\left(\frac{\delta\rho}{\rho_0}\right) =- \left(\frac{\delta\rho}{\rho_0}+1\right)\xi,
\end{equation}
\label{aeb_two}
\end{subequations}

\end{widetext}

with the following definitions,

\begin{equation}
 P_T= c_s\rho_0 {\delta \tilde p} +\frac{\rho_0}{2}(\tilde b^2+2{\bf \tilde b\cdot V_a}),
 \end{equation}
 \begin{equation}
  {\bf \tilde B=\tilde b+V_a},
\end{equation}
\begin{equation}
\frac{dR}{dt}=U,\qquad \frac{d L_x}{dt}=L_x U^\prime,
\end{equation}
\begin{equation}
\hat P_\eta=\left(\frac{2U}{R},\,\frac{U}{R}+U^\prime,\,\frac{U}{R}+U^\prime\right),
\end{equation}
\begin{equation}
\hat P_J=\left(\frac{2U}{R}+U^\prime,\,\frac{U}{R}+U^\prime,\,\frac{U}{R} +U^\prime\right),
\end{equation}
\begin{equation}
 \bn=\left(\frac{L_{0x}}{L_x}\frac{\p}{\p x},\,\frac{R_{0}}{R}\frac{\p}{\p y},\,\frac{R_{0}}{R}\frac{\p}{\p z}\right),
\end{equation}
\begin{equation}
\begin{split}
\{{\bf B,A}\}=&{\bf\hat x}\left[(\partial_y {\bf B})\cdot\bn A_z-(\partial_z{\bf B})\cdot\bn A_y \right]+\\
&{\bf\hat y}\left[(\partial_z {\bf B})\cdot\bn A_x-(\partial_x{\bf B})\cdot\bn A_z \right]+\\
&{\bf\hat z}\left[(\partial_x {\bf B})\cdot\bn A_y-(\partial_y{\bf B})\cdot\bn A_x \right].
\end{split}
\end{equation}

We defer to Appendix~\ref{2d}  the corresponding  AEB equations for the two-dimensional case. Notice that $\meta^\mp$ represent incompressible Alfv\'en modes, and that now a coefficient of the form ($1\pm V_{a,x}/U$) multiplies the time derivative of outwards (upper sign) and inwards (lower sign) modes -- in analogy with what happens in the fixed frame of reference but with respect to the radial variable. Such a coefficient, while being theoretically correct, introduces a formal, apparent singularity for inwards waves at the  Alfv\'en critical point, where $V_{a,x}=U$, that in general can be removed by imposing proper boundary conditions. Inhomogeneity indeed couples counter-propagating modes, so that an outward propagating mode $\meta^-$ forces an ``anomalous" $\meta^+$ component (which has the same phase of $\meta^-$ and hence it is not singular), plus, in general, a reflected inward propagating $\meta^+$ (that has a singularity at the critical point), see also Appendix~\ref{sing}. Since the latter is moving inwards, it will never reach the critical point and the system is not, in principle, singular. The point is that  the AEB is periodic, and the singular reflected component of $\meta^+$  propagates within the box while approaching the critical point. This evidently introduces numerical stability problems, if  the regularity condition, i.e. that at the critical point $\meta^+$ must be given only by the anomalous, forced component, is not imposed. In other words,  one must have  that both left- and right-hand-side of eq.~(\ref{meta}), lower sign, are equal to zero when $U=V_{a,x}$, and this gives a constraint on the functional form of $\meta^+$ at the critical point. In order to satisfy this condition, we have used an Euler semi-implicit method for the time advancement of $\meta^+$, in which the propagation ${\bf V_{a}\cdot \bn }\meta^+$ and the linear terms proportional to $\meta^+$ are implicit. 

This can be seen as follows: after Fourier decomposition, the evolution equation for $\meta_k^+$ takes the following schematic form, with both linear and nonlinear contributions:

\begin{equation}
\begin{split}
\left( 1-\frac{V_{a,x}}{U}\right)\frac{\p \meta_k^+}{\p t}=&[i{\bf k \cdot V_{a}}+F(R)]\meta_k^+\\
&+G(R)\meta_k^-+\text{n.l.}
\end{split}
\end{equation}

Now, the Euler semi-implicit scheme with time-step $\Delta t=t_{n+1}-t_n$,

\begin{widetext}
\begin{equation}
\left( 1-\frac{V_{a,x}}{U}\right)\frac{\meta_k^+(t_{n+1})-\meta_k^+(t_{n})}{ \Delta t}=
\left[i{\bf k \cdot V_{a}}+F(R)\right]\meta_k^+(t_{n+1})+G(R)\meta_k^-(t_n)+\text{n.l.}(t_{n}),
\end{equation}
\end{widetext}

 allows to remove the singularity and  gives the correct regularity condition on $\meta^+$ at leading order in fluctuations'  amplitude. An explicit Runge-Kutta  scheme is used for the time advancement of all other fields. 

\section{Properties of the AEB model and numerical tests}
\label{properties}

\begin{figure}[th]
\begin{center}
\includegraphics[width=0.4\textwidth]{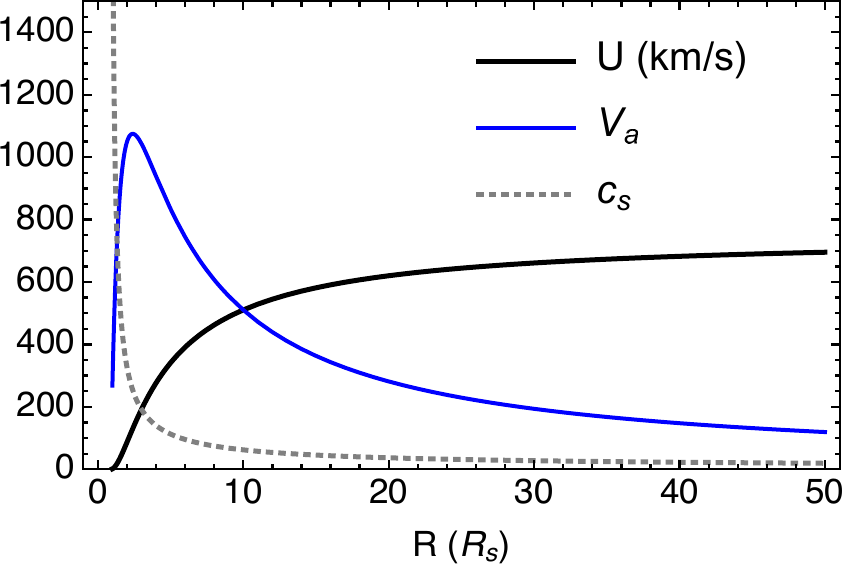}
\caption{Solar wind model for the AEB: solar wind speed profile $U$ and resulting radial evolution of the Alfv\'en ($V_a$) and sound ($c_s$) speed.}
\label{RD0}
\end{center}
\end{figure}

In this section we discuss the properties and limits of validity of the AEB  described by eqs.~(\ref{aeb_two}), and we show that it  is suitable for the description of small scale plasma dynamics from the sub-Alfv\'enic region out to the inner heliosphere, provided  the wind is supersonic. To this end we analyze the evolution of Alfv\'enic  perturbations and the general dispersion relation of MHD waves as a function of heliocentric distance. For the sake of simplicity, and because here we are mainly concerned  with the dynamics close to the  Alfv\'en critical point, we consider fluctuations within a radial magnetic field. In particular, we take a flux tube of fast solar wind with asymptotic speed $U_\infty=750$~km/s, and we assume an adiabatic closure, thus  the polytropic index is $\gamma=5/3$. In Fig.~\ref{RD0} we show the solar wind speed  $U(R)$ (solid black) and the resulting Alfv\'en (solid blue) and sound (dashed gray) speed as a function of the  heliocentric distance $R$. In our solar wind model we have placed the Alfv\'en critical point at $10\, R_s$  and the sonic point at about $3\, R_s$. In the following, we discuss  via direct analyses that shear Alfv\'en and oblique magnetosonic modes propagating within this background equilibrium are well described by the AEB model. Next we show a numerical test of an Alfv\'en wave in parallel propagation.

\subsection{Normal mode analysis}
\label{normal}

As already mentioned in the Introduction, it is known that in the case of an inhomogeneous medium with  flow the conservation  of wave energy density of a given mode of fluctuation is generalized  to the conservation law of wave action  $\mathcal E/\omega$, where $\mathcal{E}=1/2\rho_0u^2$   is the wave energy density  and $\omega$ its intrinsic frequency (frequency in the plasma rest frame), provided frequencies are larger than $\tau_e^{-1}$ (WKB approximation). Within this limit, and for a steady  background with spherical symmetry, conservation of the flux $S$ of the wave action  across a spherical surface  reads $(V_g+U)r^2\rho_0u^2/\omega=S$, where $V_g$ is the radial component of the group velocity~\citep{bretherton_1969,jacques_ApJ_1977}. Such a conservation law leads to a  radial variation of the amplitude of $u$ which, in most cases, peaks more or less where the radial group velocity crosses the flow  speed.  In particular, for outwards  Alfv\'en waves  $S=u^2(U+V_a)^2/(UV_a)$, so that their amplitude peaks  at the Alfv\'en critical point, and next it decreases as $V_a\sim r^{-1}$ for $V_a\ll U$:
\begin{equation}
u^2\sim \frac{UV_a}{(U+V_a)^2}.
\label{sa}
\end{equation}

The conservation law of wave action can be extended to the so-called non-WKB limit in which the coupling between counter-propagating modes induced by the background inhomogeneity  cannot be neglected~\citep{heinemann_JGR_1980}. This is necessary especially  for  the lower frequencies with respect to $\tau_e^{-1}$.  Although our derivation is rigorously valid only for $\omega \tau_e \gg 1$, in practice interesting results may also be derived when larger spatial and temporal scales are included in the simulations (see~\citet{grappin_JGR_1996} for the extreme case of co-rotating interaction regions). Therefore demonstrating  a greater fidelity also at such lower frequencies is important and, in particular, a correct description of inwards Alfv\'enic modes is central to any study of turbulence  in the solar wind and in particular of wave reflection driven turbulence. 

In this regard,  consider plane waves in parallel propagation and the two-dimensional AEB equations   given in Appendix~\ref{2d}.  In this case  $u_y\mp\tilde b_y=-(i/k) \eta^\mp$, where $k=k_*(L_{0x}/L_x)$, and from eqs.~(\ref{2D}) one recovers the well known  expressions for small amplitude shear  Alfv\'en waves ${\bf z}^\mp={\bf u\mp\tilde b}$ in the non-WKB limit:

\begin{widetext}
\begin{equation}
\left(1+\frac{V_a}{U}\right)\frac{\partial {{\bf z}}^{-}}{\partial t}=- ikV_a {\bf z}^{-}+ {\bf z}^+\left(\frac{V_a-U}{R}\right)+\left(\frac{\bf {z}^+-{z}^-}{2}\right)\left(\frac{\boldsymbol\nabla\cdot{\bf U}}{2}-\boldsymbol\nabla\cdot{\bf V}_a\right),
\label{aebZ}
\end{equation}
\begin{equation}
\left(1-\frac{V_a}{U}\right)\frac{\partial {\bf z}^{+}}{\partial t}= ikV_a {\bf z}^{+}-{\bf z}^-\left(\frac{V_a+U}{R}\right)+\left(\frac{\bf {z}^--{z}^+}{2}\right)\left(\frac{\boldsymbol\nabla\cdot{\bf U}}{2}+\boldsymbol\nabla\cdot{\bf V}_a\right).
\label{aebZ1}
\end{equation}
\end{widetext}


From  eqs.~(\ref{aebZ})--(\ref{aebZ1}) follows our counterpart of the generalized  equation of wave action conservation in the solar wind, where the radial dependence is now transformed into a time dependence,

\begin{widetext}
\begin{equation}
\frac{\partial }{\partial t}(|z^-|^2-|z^+|^2)=-\left[|z^-|^2\left(\frac{\boldsymbol\nabla\cdot{\bf U}}{2}+ \boldsymbol\nabla\cdot{\bf V_a}\right)- |z^+|^2\left(\frac{\boldsymbol\nabla\cdot{\bf U}}{2}- \boldsymbol\nabla\cdot{\bf V_a}\right)\right].
\label{invariant}
\end{equation}
\end{widetext}

After some algebra, ~eq.~(\ref{invariant}) leads to the known conservation law of the total wave action flux $S^*$, 
\begin{equation}
S^*= |z^-|^2\left[\frac{(U+V_a)^2}{UV_a} \right]- |z^+|^2\left[\frac{(U-V_a)^2}{UV_a} \right].
 \label{wa}
\end{equation}

As can be seen, the large scale gradient and the Alfv\'en speed advection terms, in addition to the standard expansion ones, are necessary for the description of the evolution of energy densities close to the Alfv\'en critical point. In particular, the term $(1\pm V_a/U)$ which multiplies the time derivative ensures the correct radial evolution of the intrinsic frequencies. 

Indeed, the evolution of energy densities with heliocentric distance crucially depends on the  radial dependence of the intrinsic frequency $\omega$. The dispersion relation of Alfv\'en waves in the AEB can be easily obtained from eqs.~(\ref{aebZ})--(\ref{aebZ1}) to leading order in $(k_*R_0)^{-1}$, assuming a solution of the form $A(t)\exp[-i\int^t{\omega (t^\prime)dt^\prime}+ik_*x]$. This is  a standard procedure but a detailed calculation can be found also in Appendix~\ref{sing}. We obtain in this way 

\begin{equation}
\begin{split}
\omega(R)&=\pm k_*\frac{L_{0x}}{L_x}\frac{UV_a}{U\pm V_a}\\
&=\pm\Omega\frac{V_a}{U\pm V_a}\\
&=\pm k_{abs}(R) V_a,
\end{split}
\end{equation} 

where $\Omega=kU$ is the (constant) frequency in the fixed frame of reference. In this regard, recall that, because of the stretching of the box, the derivative in the radial ($x$) direction is rescaled, and as a consequence the radial wave vector is $k=k_*(L_{0x}/L_x)$. It can be easily verified that $k U=const.$, and that it corresponds to the absolute frequency $\Omega$ when $V_a/U\ll 1$.  Also, $k_{abs}$ can be recognized as the wave vector of Alfv\'en waves in the fixed frame, and it  must be considered as the effective wave vector within the AEB, being the one consistent with the intrinsic dispersion relation.  If the term $\pm V_a/U$  that multiplies the time derivative was neglected, we would not have the correct radial evolution of $\omega(R)$ close to the critical point. This is  what happens for outwards and inwards radial sound waves (represented respectively by $y^\pm=u_x\pm\tilde p$) at the sonic point, for which the same considerations as those for Alfv\'en waves hold. 

\begin{figure}[th]
\begin{center}
\includegraphics[width=0.4\textwidth]{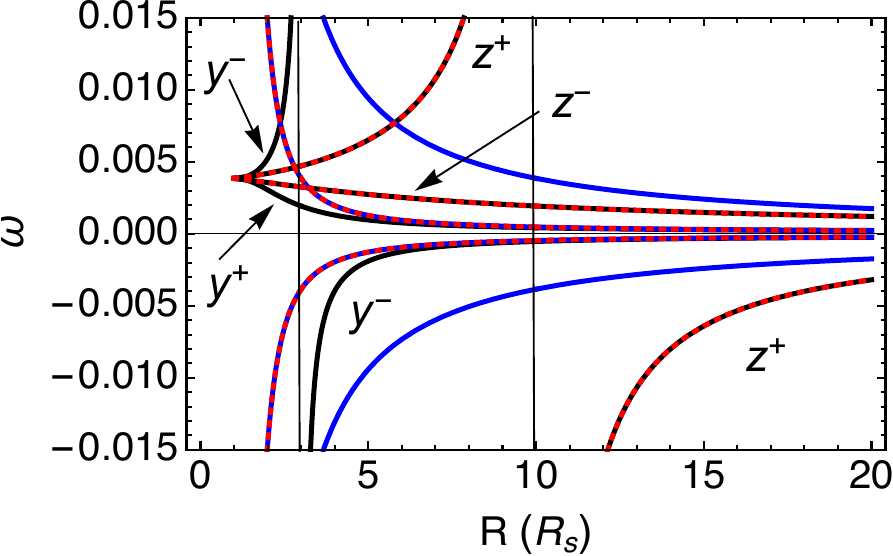}
\includegraphics[width=0.4\textwidth]{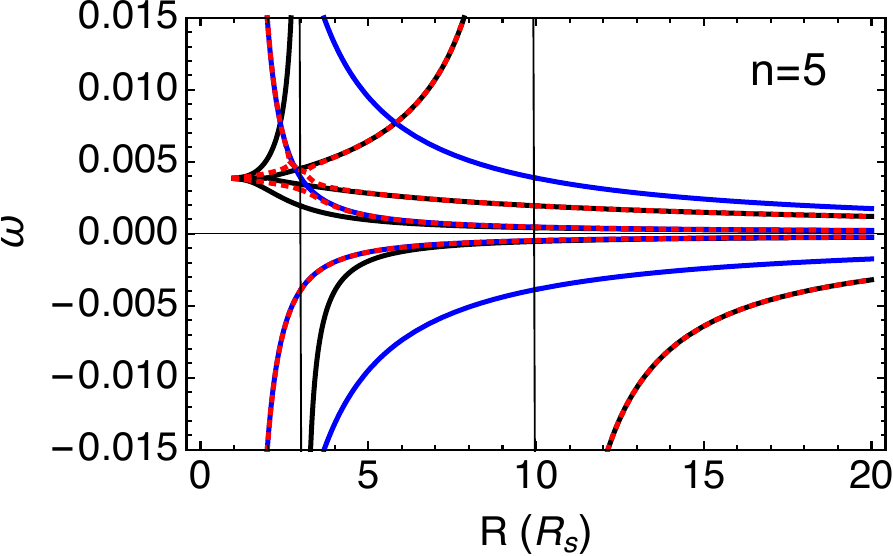}
\includegraphics[width=0.4\textwidth]{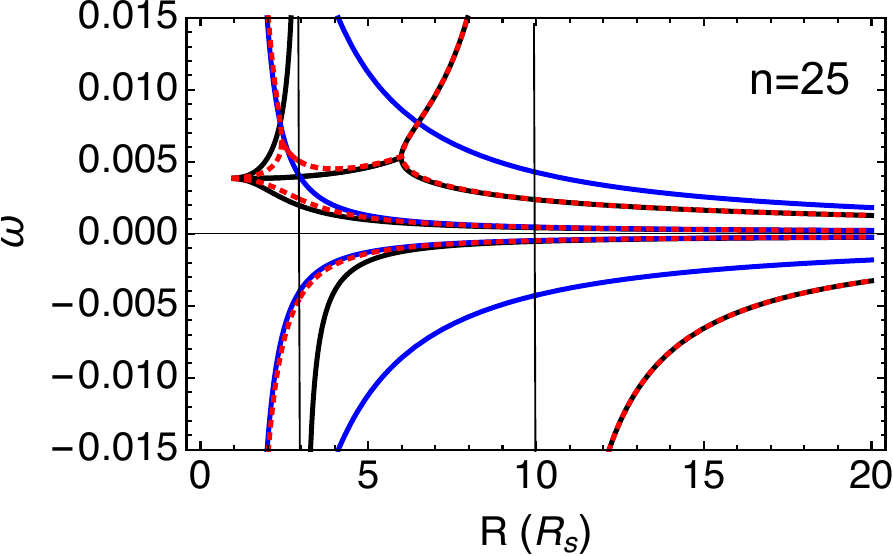}
\caption{Intrinsic frequency $\omega$ (Rad/s) vs. heliocentric distance for $\Omega=0.004$~Rad/s. Upper panel: Alfv\'en ($z^\mp$) and sound ($y^\pm$) waves. Middle and bottom panels: fast and slow modes in oblique propagation for two different values of perpendicular wave number. Black lines correspond to the theoretical solution,  red-dashed lines to the solution in the AEB, and  blue lines to the AEB solution without the large scale advection at $V_a$.}
\label{RD1}
\end{center}
\end{figure}

To conclude this first part, we show that, as desired, the AEB reproduces the intrinsic frequencies of MHD waves  to an excellent approximation, provided $U>c_s$. In Fig.~\ref{RD1}  we plot the radial evolution of  $\omega(R)$ for  the shear Alfv\'en and parallel sound waves (upper panel) and of fast and slow mode waves  propagating in the $\{x,y\}$ plane at two different angles (middle and bottom panels). In these plots we compare $\omega(R)$, obtained from the AEB linearized equations, with the theoretical solution, that we label  $\omega_{abs}(R)$. The latter is the Doppler shifted frequency in the fixed frame of reference, $\omega_{abs}(R)=\Omega-k_{r,abs}U$, and $k_{r,abs}$ can be found as a function of $R$ from the MHD dispersion relation, using our solar wind model. We compare frequencies at the same values of  $k_x U=\Omega$ and of perpendicular wave number $k_y=n/R$: the black line corresponds to $\omega_{abs}$ (the exact solution), the red-dotted line corresponds to $\omega(R)$ in the AEB, and the blue-dotted line corresponds to the intrinsic frequency in the AEB if the large scale advection correction, $\pm (V_a/U)\p/\p t$, is neglected. As can be seen, even for relatively large angles the AEB provides an excellent description of frequencies, provided the wind is supersonic, and it even captures the reflection point of fast oblique inwards modes (when the two branches of out- and inwards modes connect). On the contrary,  Alfv\'en and fast modes would be poorly approximated in our regions of interest without the large scale advection term.

\subsection{Numerical test: Alfv\'en wave propagation across the  critical point}
We consider the same fast solar wind flux tube with radial magnetic field shown in Fig.~\ref{RD0}, from an initial  heliocentric distance $R_0=6\, R_s$. The initial values of the background speeds are $U_{*}=390$~km/s, $V_{*a}=750$~km/s and $c_{*s}=100$~km/s. The inverse expansion time  is $\tau_e^{-1}=U_\infty/R_0\simeq 2\times10^{-4}$~s$^{-1}$.  We consider  a small amplitude wave packet polarized in the $z$ direction with initial conditions 

\begin{equation}
\begin{split}
&{ z^-}(x,0)=A\exp\left\{-[(x-x_0)/\Delta]^2\right\}\cos(k_*x), \\
& z^+(x,0)=0.
\end{split}
\end{equation}

The amplitude is set to $A/V_{*a}=0.001$, the initial wavelength is $\lambda_*=0.75\,R_s$, that corresponds to a frequency $\Omega=0.005$~Rad/s, and the width of the envelop is $\Delta=1\, R_s$. We choose a localized wave packet instead of a monochromatic wave for illustrative purposes, as this allows to better show the generation of the two components of the inward mode,  the forced  and the reflected one, and to check their  evolution.

\begin{figure}[htbp]
\begin{center}
\includegraphics[width=0.4\textwidth]{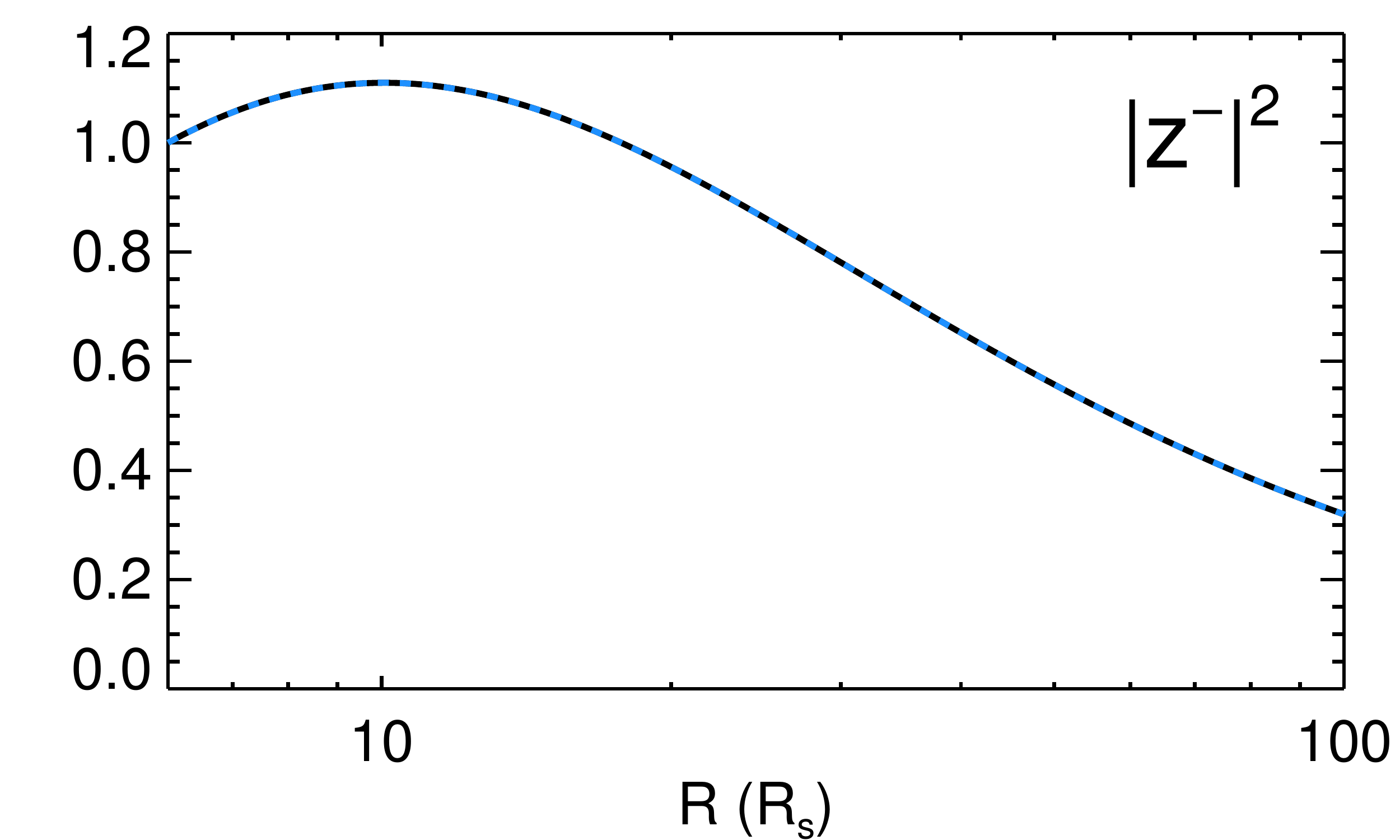}\quad
\includegraphics[width=0.4\textwidth]{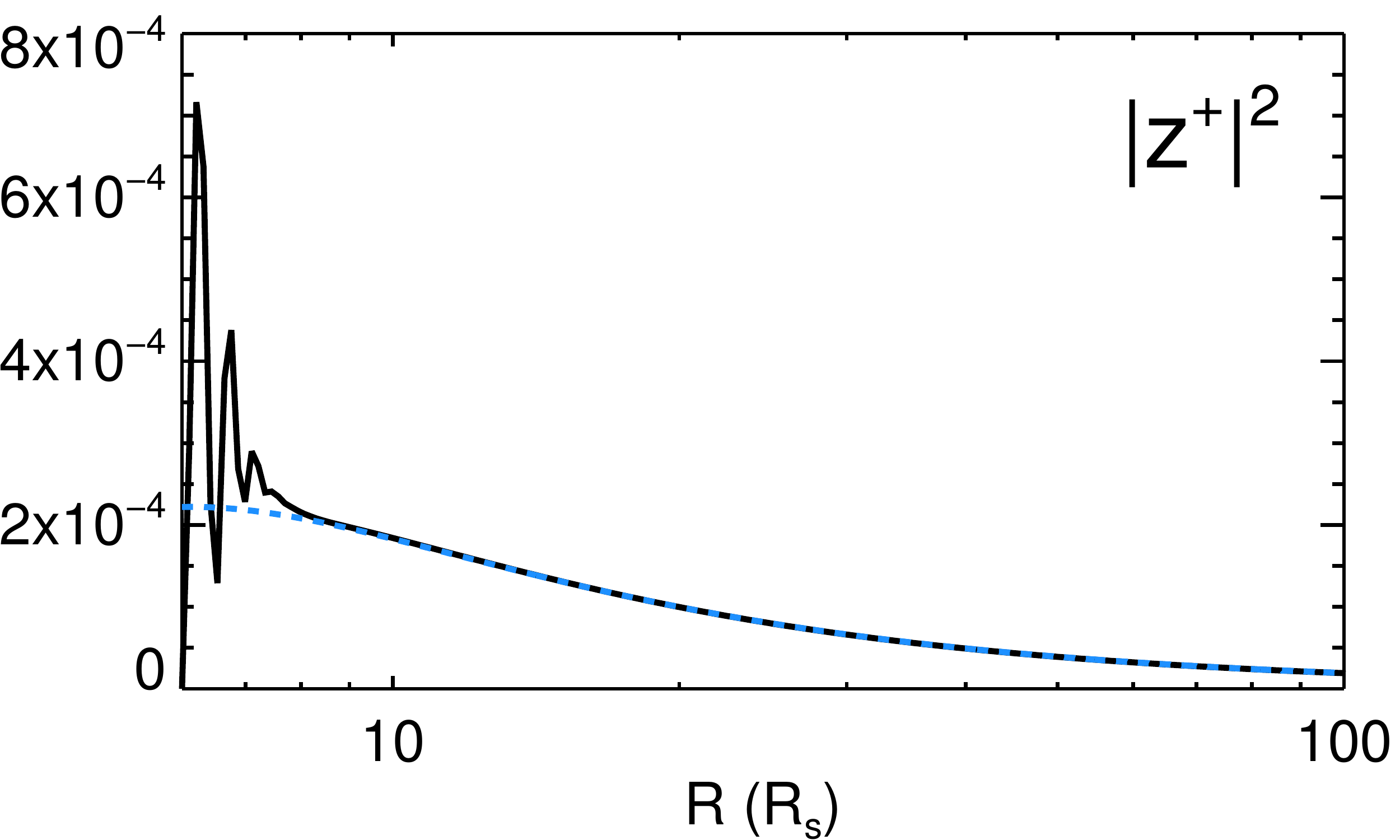}
\caption{Radial evolution of $|z^-|^2$ (upper panel) and  $|z^+|^2$ (bottom panel), normalized at the initial value $|z^-(0)|^2$. The dotted line represents the exact solution obtained by solving numerically the  boundary value problem of a stationary Alfv\'en wave in the Eulerian frame.}
\label{alfven}
\end{center}
\end{figure}

In Fig.~\ref{alfven} we show the  evolution of the root-mean-square amplitude of ${ z}^-$ and $z^+$ as a function of the average heliocentric distance $R$. We compare our numerical solution with that  obtained by numerically solving  the  boundary value problem of a stationary, monochromatic Alfv\'en wave in the Eulerian frame at the same $\Omega$, shown as a dotted line. As can be seen in the upper panel, our results for $z^-$ are in agreement with theory, and the amplitude is given to an excellent approximation by $|z^-|^2\propto UV_a/(U+V_a)^2$. The envelop of $|z^+|^2$ also evolves following the theoretical solution (bottom panel). The presence of rapid oscillations at the beginning is due to  beats between the forced  and the reflected components of $z^+$. This stems from the fact that the AEB solves an initial value problem. Therefore, if at $t=0$ only $z^-$ is given,  then a forced and a reflected $z^+$ are generated accordingly, causing beats when they superpose. In Appendix~\ref{sing} we derive the analytical solution of the wave equations~(\ref{aebZ})--(\ref{aebZ1}) via a two-scale analysis that illustrates this point. The numerical scheme however behaves like an effective filter and it damps the singular reflected component while approaching the critical point, at $10\, R_s$. In this way the reflected mode does not propagate beyond the critical point, as expected,  preserving the causal disconnection between the sub-Alfv\'enic  and the super-Alfv\'enic corona.

\begin{figure}[htbp]
\begin{center}
\includegraphics[width=0.45\textwidth]{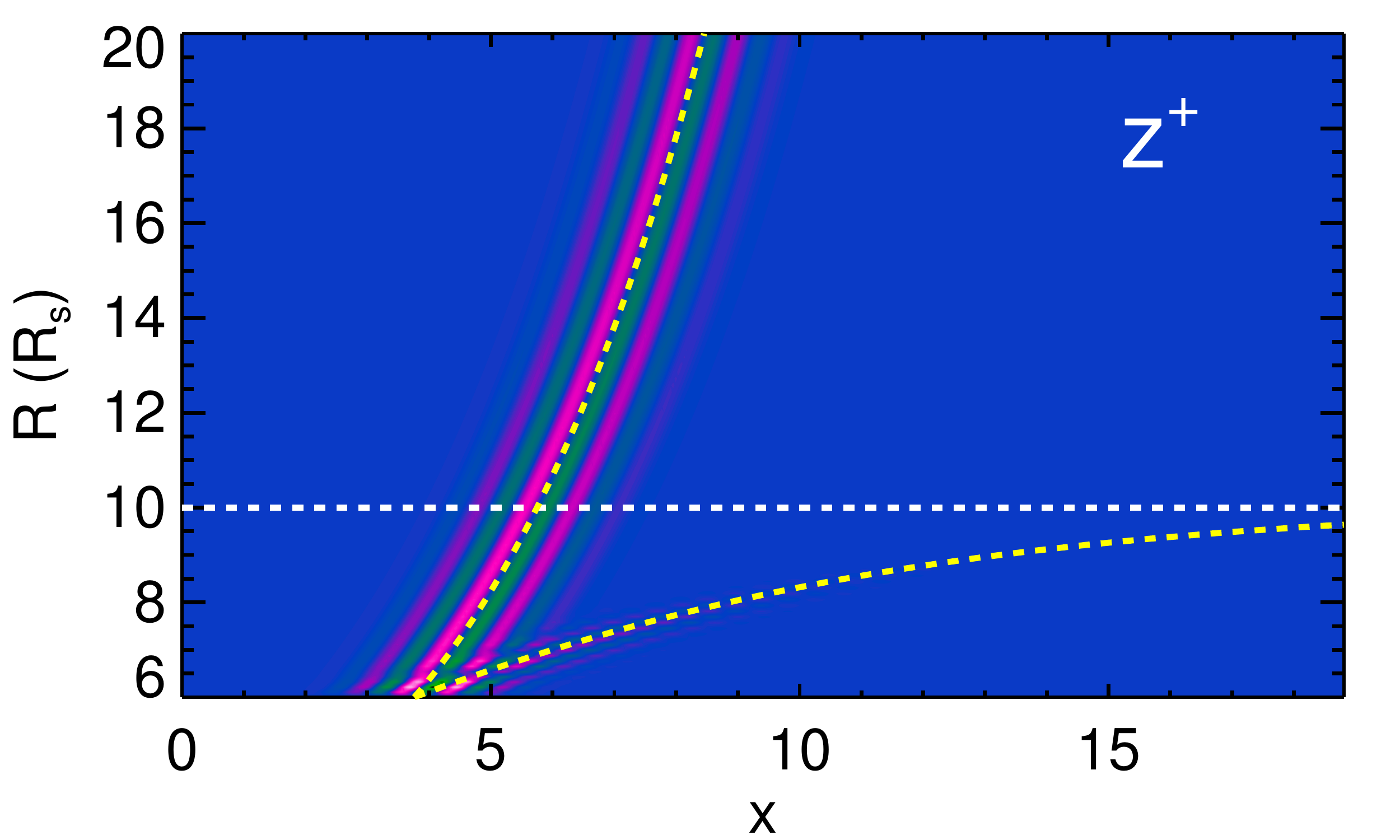}
\caption{Contours of $z^+$: the white line indicates the critical point and the yellow-dashed lines represent the characteristic $x(R)$ of the anomalous forced component and of the reflected singular component given in eq.~(\ref{char-})--(\ref{char+}). }
\label{cont}
\end{center}
\end{figure}

This can be seen also by inspecting Fig.~\ref{cont}, where the contours of $z^+$ in $\{x,R\}$ space are displayed. One can  see that there are two components  propagating at two different phase speeds that decouple with time (heliocentric distance). These two components are tracked by the yellow-dotted lines  corresponding to the characteristics of the system: the forced component of $z^+$, by definition, follows the characteristic of $z^-$ given by

\begin{equation}
\frac{dx}{dR}=\frac{V_a}{U+V_a} \frac{L_{0x}}{L_x},
\label{char-}
\end{equation}

while the characteristic of the reflected component is given by

\begin{equation}
\frac{dx}{dR}=-\frac{V_a}{U-V_a} \frac{L_{0x}}{L_x}.
\label{char+}
\end{equation}

\begin{figure}[htb]
\begin{center}
\includegraphics[width=0.45\textwidth]{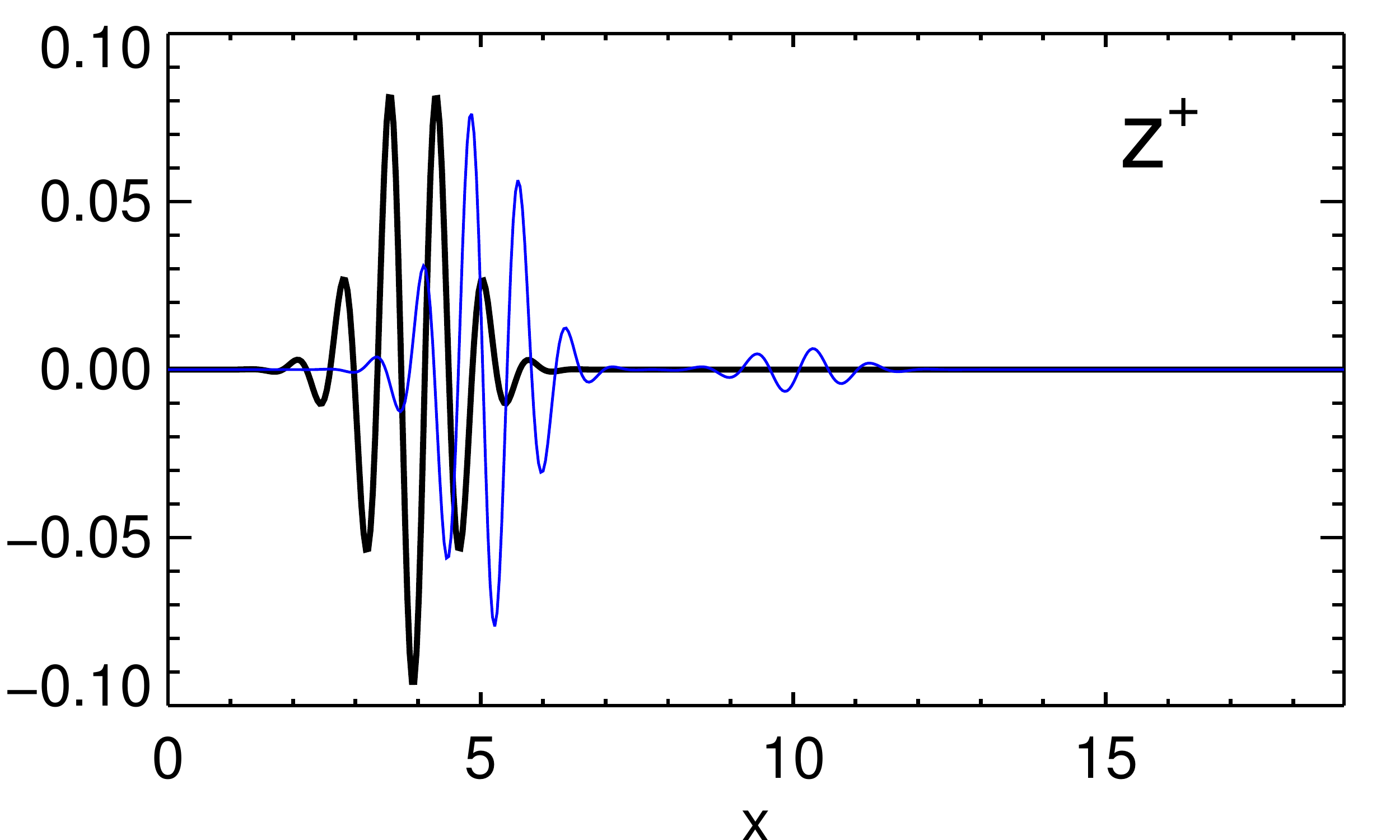}
\caption{Wave form of $z^+$, with amplitude normalized to $|z^-(0)|$, at two different times inside the Alfv\'en point: the  black line corresponds to time $t\Omega=0.9$ while the thin blue one to $t\Omega=17$, the latter showing the forced component moving with $z^-$ (large amplitude component) and the reflected one (small amplitude component).}
\label{alfven2}
\end{center}
\end{figure}

In Fig.~\ref{alfven2} we show for reference the waveform of $z^+$ at $t\Omega=0.9$ ($R=6.1\,R_s$, black line) and at $t\Omega=17$ ($R=8.3\,R_s$, blue thin line) that shows the separation of the two components of ${z^+}$ during the evolution of the initial wave packet, and the damping of the singular component.

%
%

\section{Summary and conclusions}
\label{discussion}
We have presented the accelerating expanding box that extends and improves the previous expanding box model in two directions. The original EBM is suitable for describing plasma dynamics in the co-moving frame of the solar wind  when a uniform flow is assumed at large distances from the Sun, i.e.,  when the flow is both supersonic and super-alfv\'enic, $U\gg c_s$ and $U\gg V_a$. The present model is an extension and improvement in  that:

$i)$ We allow for a non uniformity of the solar wind speed at first order inside the box (i.e. we retain up to first derivative of $U(R)$). This causes an additional stretching of the box in the $x$ (radial) direction and additional terms $\propto U^\prime$  coming from the ${\boldsymbol\nabla\cdot {\bf U}}$ and from the  material derivative of ${\bf U}$. These terms cause the average pressure, density and azimuthal magnetic field to decrease with heliocentric distance faster than in the EBM.

$ii)$ In order to be consistent with the description of the dynamics close and across the Alfv\'en critical point,  we retain terms of order $V_a/U$ (neglecting effects of the Parker spiral close to the Sun), which are not taken into account in the standard EBM.  These new terms modify the decay rates of fluctuations only, and introduce additional reflection terms that allow for the conservation of the adiabatic invariants of Alfv\'en waves, i.e. the generalized wave action flux, given in eq.~(\ref{wa}).  

To summarize, the  AEB model is described by eqs.~(\ref{aeb1}), that have been re-written in terms of generalized Els\"asser variables in eqs.~(\ref{aeb_two}). The latter formulation has the advantage of separating the compressible and incompressible parts of the velocity, as well as  decoupling forward and backward Alfv\'enic modes, necessary for the numerical integration of the equations across the  Alfv\'en critical point. As we have shown,  the model equations properly describe the  Alfv\'en critical point crossing without unphysical accumulation of reflected modes at the boundary between sub- and super-Aflv\'enic corona.  Finally, linear analysis of the normal modes of the AEB shows that the theoretical evolution of the intrinsic frequencies of MHD waves is recovered in the supersonic wind.   

\acknowledgements A.T. thanks D. Del Sarto for his useful comments and discussions. This work was supported by the NASA program LWS, grant NNX13AF81G as well as the NASA Solar Probe Plus Observatory Scientist grant. This work used the Extreme Science and Engineering Discovery Environment (XSEDE), which is supported by National Science Foundation grant number ACI-1053575.


\appendix

\section{Large scale gradients in the induction equation}
\label{LSC_induction}

In this Appendix we discuss  the approximations done to derive the large scale gradients in the induction equation, and we show how to obtain eq.~(\ref{ind_inc}) from eq.~(\ref{ind}) in a form consistent with $\bn\cdot {\bf b}=0$. We therefore consider only  ${\bf \bn\times u\times B_0}$, since both the nonlinear  (${\bf \bn\times u\times b}$) and the expansion  (${\bf \bn\times U\times b}$) terms do not affect the conservation of the magnetic field divergence. 

We  develop to first order in  $\epsilon$ and $\alpha$ vectors,  derivatives and  average magnetic field, so as to obtain

\begin{equation}
{\bf\hat{r}}= {\bf\hat{x}} +\frac{y}{R}{\bf\hat{y}}+\frac{z}{R}{\bf\hat{z}},\quad {\bf \hat{\boldsymbol\theta}}=\frac{z}{R}{\bf\hat{x}}- {\bf\hat{z}} ,\quad {\bf \hat{\boldsymbol\varphi}}={\bf\hat{y}}  -\frac{y}{R}{\bf\hat{x}},
\label{b0} 
\end{equation}
\begin{equation}
\left\{\frac{\partial}{ \partial r},\,\frac{1}{r\sin\theta}\frac{\partial}{\partial \varphi},\,\frac{\partial}{r\partial\theta} \right\}=\left\{ \partial_x+ \frac{y}{R}\partial_y+\frac{z}{R}\partial_z, -\frac{y}{R}\partial_x+\partial_y,\,\frac{z}{R}\partial_x-\partial_z\right\},
\label{expan_3}
\end{equation} 
\begin{equation}
B_{0,r}(r)=\bar B_{0,x}(R)\left(  1-\frac{2\delta x}{R} \right),\qquad B_{0,\varphi}(r)=\bar B_{0,y}(R)\left( 1-\frac{\delta x}{R} - \frac{U^\prime}{U}\delta x \right),
\label{b01} 
\end{equation}

where

\begin{equation}
\bar B_{0,x}(R)\equiv B_{*,x}\frac{R_0^2}{R^2},\quad \bar B_{0,y}(R)\equiv B_{*,y}\frac{R_0}{R}.
\label{b02} 
\end{equation}

 In writing eq.~(\ref{b01}) we have  expanded to linear order ${\bf B_0}(r)={\bf B_0}(R)+{\bf B_0}^\prime(R)\delta x$, and  ${\bf B_0}^\prime(R)$ is expressed by making use of the following evolution equations of magnetic field in the solar wind,
 
\begin{equation}
\frac{\partial}{\partial r}B_{0r}=-\frac{2B_{0r}}{r}
,\quad\frac{\partial}{\partial r}B_{0\varphi}=-B_{0,\varphi}\left(\frac{1}{r}+\frac{U^\prime(r)}{U(r)}\right).
\end{equation}
 
The resulting  components of ${\bf \bn\times u\times B_0}$  are:

\begin{equation}
\begin{split}
&{\bf \hat{r}}[ -{\bf B_0(\boldsymbol\nabla\cdot u)}+ {\bf B_0\cdot\boldsymbol\nabla u}-{\bf u\cdot\boldsymbol\nabla B_0}]_{\hat r}={\bf \hat{r}}\left[-B_{0,r}\left({\bf \bn\cdot u}\right)+{\bf B_0\cdot\boldsymbol\nabla}u_r+\frac{2 B_{0,r}u_r}{r}\right]=\\
&= \left[  {\bf\hat{x}} +\frac{y}{R}{\bf\hat{y}}+\frac{z}{R}{\bf\hat{z}}\right]\bar B_{0,x}(R) \left[-\left( 1-\frac{2\delta x}{R} \right)\left({\bf \bn\cdot u}\right)+\frac{2 u_x}{R}\right]\\
&+\left[  {\bf\hat{x}} +\frac{y}{R}{\bf\hat{y}}+\frac{z}{R}{\bf\hat{z}}\right]\bar B_{0,x}(R)\left[  \left( 1-\frac{2\delta x}{R} \right)\frac{\partial u_x}{\partial x} +\frac{y}{R}\left( \frac{\partial u_y}{\partial x}+ \frac{\partial u_x}{\partial y} \right) +\frac{z}{R}\left( \frac{\partial u_z}{\partial x}+ \frac{\partial u_x}{\partial z} \right)\right]\\
&+\left[  {\bf\hat{x}} +\frac{y}{R}{\bf\hat{y}}+\frac{z}{R}{\bf\hat{z}}\right]\bar B_{0,y}(R)\left[ \left( 1-\frac{\delta x}{R} - \frac{U^\prime}{U}\delta x \right)\frac{\partial u_x}{\partial y} -\frac{y}{R}\left( \frac{\partial u_x}{\partial x}- \frac{\partial u_y}{\partial y} \right) +\frac{u_y}{R}+\frac{z}{R} \frac{\partial u_z}{\partial y}\right],
\end{split}
\label{faraday_1}
\end{equation}

\begin{equation}
\begin{split}
&{\bf \hat{\boldsymbol\theta}}[ -{\bf B_0(\boldsymbol\nabla\cdot u)}+ {\bf B_0\cdot\boldsymbol\nabla u}-{\bf u\cdot\boldsymbol\nabla B_0}]_{\hat \theta}={\bf \hat{\boldsymbol\theta}}\left[{\bf B_0\cdot\boldsymbol\nabla}u_\theta-\frac{B_{0r}u_\theta}{r} \right]=\\
&=\left[  \frac{z}{R}{\bf\hat{x}} - {\bf\hat{z}} \right]\bar B_{0,x}(R) \left[ -\left( 1-\frac{2\delta x}{R} \right)\frac{\partial u_z}{\partial x} -\frac{y}{R}\frac{\partial u_z}{\partial y}+\frac{z}{R}\left( \frac{\partial u_x}{\partial x}-\frac{\partial u_z}{\partial z} \right) \right] \\
&+\left[  \frac{z}{R}{\bf\hat{x}}- {\bf\hat{z}} \right]\bar B_{0,y}(R)\left[ -\left( 1-\frac{\delta x}{R} - \frac{U^\prime}{U}\delta x \right) \frac{\partial u_z}{\partial y} +\frac{y}{R}\frac{\partial u_z}{\partial x} +\frac{z}{R}\frac{\partial u_x}{\partial y} \right]-{\bf\hat{z}}\frac{\bar B_{0,x}(R)}{R}u_z,
\end{split}
\label{faraday_2}
\end{equation}

\begin{equation}
\begin{split}
&{\bf \hat{\boldsymbol\varphi}}[ -{\bf B_0(\boldsymbol\nabla\cdot u)}+ {\bf B_0\cdot\boldsymbol\nabla u}-{\bf u\cdot\boldsymbol\nabla B_0}]_{\hat \varphi}=\\
&{\bf \hat{\boldsymbol\varphi}}\left[ -B_{0\varphi}\left(\bn\cdot {\bf u}\right)+{\bf B_0\cdot\boldsymbol\nabla}u_\varphi +\frac{B_{0\varphi}u_r}{r}+\frac{\cot\theta B_{0\varphi}u_\theta}{r} +u_rB_{0\varphi}\left(\frac{1}{r}+\frac{U^\prime}{U}\right) - \frac{u_\varphi B_{0r}}{r}  \right]=\\
&=\left[ {\bf\hat{y}}  -\frac{y}{R}{\bf\hat{x}}  \right]\bar B_{0,y}(R)\left[\left( -1+\frac{\delta x}{R} + \frac{U^\prime}{U}{\delta x} \right)\left(\bn\cdot {\bf u}\right)+\frac{2 u_x}{R}+u_x\frac{U^\prime}{U}\right]\\
&+\left[ {\bf\hat{y}}  -\frac{y}{R}{\bf\hat{x}}  \right]\bar B_{0,x}(R) \left( 1-\frac{2\delta x}{R} \right)\left[ \frac{\partial u_y}{\partial x}+\frac{z}{R}\frac{\partial u_y}{\partial z}+\frac{y}{R}\left ( \frac{\partial u_y}{\partial y}-\frac{\partial u_x}{\partial x} \right) \right]\\
&+\left[ {\bf\hat{y}}  -\frac{y}{R}{\bf\hat{x}}  \right]\bar B_{0,y}(R)\left( 1-\frac{\delta x}{R} - \frac{U^\prime}{U}{\delta x} \right)\left[ \frac{\partial u_y}{\partial y} -\frac{u_x}{R}-\frac{y}{R} \left(  \frac{\partial u_y}{\partial x}+\frac{\partial u_x}{\partial y} \right) \right]-{\bf\hat{y}} \frac{\bar B_{0,x}(R)}{R} u_y.
\end{split}
\label{faraday_3}
\end{equation}

After collecting eqs.~(\ref{faraday_1})--(\ref{faraday_3}) and neglecting the remaining higher order terms, we obtain the following  expressions for ${\bf \bn\times u\times B_0}$ in cartesian coordinates:

\begin{equation}
\begin{split}
{\bf\hat x}:-&\left[\bar B_{0,x}(R)\left( 1-2\frac{\delta x}{R}  \right)-\frac{y}{R}\bar B_{0,y}(R)\right]{\bf\boldsymbol\nabla\cdot u}+\bar B_{0,x}(R)\left [ \left( 1-\frac{2\delta x}{R}\right) \frac{\partial u_x}{\partial x} +\frac{y}{R}\frac{\partial u_x}{\partial y}+  \frac{z}{R}\frac{\partial u_x}{\partial z} \right]\\
&+\bar B_{0,y}(R)\left[\left( 1-\frac{\delta x}{R} - \frac{U^\prime}{U}{\delta x} \right)\frac{\partial u_x}{\partial y}-\frac{y}{R}\frac{\p u_x}{\p x}\right]+\frac{2 u_x\bar B_{0,x}(R)}{R},
\end{split}
\label{faraday_lin_x}
\end{equation}

\begin{equation}
\begin{split}
{\bf\hat y}: -&\left[\bar B_{0,y}(R)\left( 1-\frac{\delta x}{R} - \frac{U^\prime}{U}{\delta x} \right)+\frac{y}{R}\bar B_{0,x}(R)\right]{\bf\boldsymbol\nabla\cdot u}+\bar B_{0,x}(R)\left [ \left( 1-\frac{2\delta x}{R}\right) \frac{\partial u_y}{\partial x} +\frac{y}{R}\frac{\partial u_y}{\partial y}+  \frac{z}{R}\frac{\partial u_y}{\partial z} \right] \\
&+\bar B_{0,y}(R)\left[\left( 1-\frac{\delta x}{R} - \frac{U^\prime}{U}{\delta x} \right)\frac{\partial u_y}{\partial y}-\frac{y}{R}\frac{\p u_y}{\p x}+\frac{u_x}{R}\right]-\frac{\bar B_{0,x}(R)u_y}{R}+\bar B_{0,y}(R) u_x\left(\frac{1}{R}+\frac{U^\prime}{U}\right),
\end{split}
\label{faraday_lin_y}
\end{equation}

\begin{equation}
\begin{split}
{\bf\hat z}: -&\frac{z}{R}\bar B_{0,x}(R){\bf\boldsymbol\nabla\cdot u}+\bar B_{0,x}(R)\left [ \left( 1-\frac{2\delta x}{R}\right) \frac{\partial u_z}{\partial x} +\frac{y}{R}\frac{\partial u_z}{\partial y}+  \frac{z}{R}\frac{\partial u_z}{\partial z} \right]\\
&+\bar B_{0,y}(R)\left[ \left( 1-\frac{\delta x}{R} - \frac{U^\prime}{U}{\delta x} \right) \frac{\p u_z}{\p y}-\frac{y}{R}\frac{\p u_z}{\p x}\right]-\frac{\bar B_{0,x}(R)u_z}{R}.
\end{split}
\label{faraday_lin_z}
\end{equation}

The  reduced components of ${\bf \bn\times u\times B_0}$ given in~(\ref{faraday_lin_x})-(\ref{faraday_lin_z}) have zero divergence, also in the simpler case with $\bar B_{0,y}=0$ that is considered specifically in this paper. It is also easy to recognize the origin of each term: the first one represents ${\bf B_0}({\bf \boldsymbol\nabla\cdot u})$; the second and third terms correspond to the advection at the Alfv\'en velocity ${\bf B_0\cdot\boldsymbol\nabla u}$; the last ones, of the form $\sim \bar B_{0} u/R$, are the large scale gradients of the magnetic field originating from ${\bf u\cdot\boldsymbol\nabla B_0}$. Although the decomposition given in eqs.~(\ref{faraday_lin_x})-(\ref{faraday_lin_z}) is rigorous to our order of approximation, there is now an explicit dependence on local coordinates, whereas we want to recover local homogeneity and periodicity. We therefore neglect first order geometrical  terms $\propto y/R,\,z/R$, the local inhomogeneity of ${\bf B_0}$, which is $\propto\delta x/R$, and  $2u_x/R$ (in any case this term does not contribute to linear Alfv\'en waves), while we retain the large scale gradients in the perpendicular directions and  advection:

\begin{equation}
 -{\bf B_0(\boldsymbol\nabla\cdot u)}+ {\bf B_0\cdot\boldsymbol\nabla u} -{\bf u\cdot\boldsymbol\nabla B_0} \simeq-{\bf \bar B_{0}}( {\bf{\boldsymbol\nabla}\cdot u} ) + ({\bf \bar B_0\cdot{\boldsymbol\nabla}}){\bf  u}+\bar B_{0,x}\frac{\p {\bf u}}{\p R} -\frac{{\bf u}_\bot\bar B_{0,x}}{R}.
 \label{eq:ap}
\end{equation}

The  expression for the normalized magnetic field consistent with eq.~(\ref{eq:ap}) is

\begin{equation}
 -{\bf V_a(\boldsymbol\nabla\cdot u)}+ {\bf V_a\cdot\boldsymbol\nabla u} -{\bf u\cdot\boldsymbol\nabla V_a}-\frac{1}{2}{\bf V_a}u_r(\ln\rho_0)^\prime \simeq-{\bf \bar V_a}( {\bf{\boldsymbol\nabla}\cdot u} ) + {\bf \bar V_a\cdot{\boldsymbol\nabla}}{\bf  u}+\bar V_{a,x}\frac{\p {\bf u}}{\p R} -\frac{{\bf u}_\bot\bar V_{a,x}}{R}.
 \end{equation}
 
Change to the co-moving frame is now straightforward with the transformation

\begin{equation}
\boldsymbol\nabla=\frac{L_0}{L}\frac{\partial}{\partial{\bar x}}{\bf\hat x}+ \frac{R_0}{R(t)}\frac{\partial}{\partial{\bar y}}{\bf\hat y}+ \frac{R_0}{R(t)}\frac{\partial}{\partial{\bar z}}{\bf\hat z}\equiv\bar{\boldsymbol\nabla},\quad \frac{\p}{\p R}=\frac{1}{U}\frac{\p}{\p t},
 \label{faraday_modif0}
\end{equation}

which allows us to write
\begin{equation}
 -{\bf V_a(\boldsymbol\nabla\cdot u)}+ {\bf V_a\cdot\boldsymbol\nabla u} -{\bf u\cdot\boldsymbol\nabla V_a} \simeq-{\bf \bar V_{a}}( {\bf\bar{\boldsymbol\nabla}\cdot\bar u} ) + ({\bf \bar V_a\cdot\bar{\boldsymbol\nabla}}){\bf \bar u} -\frac{{\bf\bar u}_\bot\bar V_{a,x}}{R}+\frac{V_{a,x}}{U}\frac{\p {\bf \bar u}}{\p t}.
 \label{faraday_modif}
\end{equation}

The modified induction equation as given by~(\ref{faraday_modif})  does not conserve the divergence of the magnetic field because  the last two terms depend on the velocity ${\bf\bar u}$. In order to conserve the divergence-free condition, we have to impose that such couplings involve the incompressible  part of the velocity ${\bf \bar{u}}^{\text {inc}}$, as indeed is the case for Alfv\'en waves. Notice that our expanding box coordinates are time-dependent, so that the divergence of eq.~(\ref{faraday_modif}) leaves a residual term $U^\prime/U \bar V_{a,x}\p \bar u_x/\p \bar x$ that comes from the time derivative of coordinates. Although for small amplitude Alfv\'en waves this equals zero (in our approximation of radial magnetic field close to the Sun), we formally balance it by adding  $U^\prime/U \bar V_{a,x} u_x$ along the ${\bf \hat x}$ direction.

\section{Two-dimensional model equations}
\label{2d}

For the two-dimensional case, velocity and magnetic field can be decomposed via two scalar potentials $\chi$ and $\phi$ plus a component along~${\bf \hat z}$:
\begin{equation}
{\bf u=\bn\times\phi\hat z}+\bn\chi+u_z{\bf \hat z},\quad -\nabla^2 \phi=\eta,
\end{equation}
\begin{equation}
{\bf b=\bn\times\psi\hat z}+b_z{\bf \hat z },\quad -\nabla^2 \psi=j.
\end{equation}
In this case we integrate the variables $\eta^\mp=\eta\mp\tilde j$ and $z^\mp=u_z\mp\tilde b_z$, with $\tilde j=j/\sqrt{4\pi\rho_0}$ and $\tilde b_z=b_z/\sqrt{4\pi\rho_0}$:

\begin{subequations}
\begin{equation}
\begin{split}
\left(1\pm\frac{V_{a,x}}{U}\right)\frac{\p { \eta^\mp}}{\p t}&+{\bf u\cdot\bn\eta^\mp}=\mp{\bf \tilde B}\cdot\bn \eta^\mp\pm\frac{V_{a,r}}{R} \eta^\pm-\frac{1}{2}\frac{ \eta^+- \eta^-}{2}\left(\frac{2}{R}+\frac{U^\prime}{U}\right)(V_{a,r}\pm U)\\
& \mp\{{\bf\tilde b,u\}}_z \pm\{{\bf u, \tilde b\}}_z +\frac{\rho_0}{\rho} \{{\bf \tilde b,\tilde b\}}_z-\left(U^\prime+\frac{U}{R}  \right)  \frac{\eta^++ \eta^-}{2}\pm \left(U^\prime+\frac{U}{R}  \right) \frac{\eta^+- \eta^-}{2}\\
&-\frac{\delta}{1+\delta}{\bf \tilde B}\cdot\bn \tilde j-\left(\frac{\rho_0}{\rho^2}\bn\delta\rho\times({\bf\tilde B\cdot\bn\tilde b})\pm\bn\xi\times{\bf\tilde B}+\frac{1}{\rho^2}\bn\delta\rho\times\bn P_T\right)_z-\xi\eta^\mp\\
&+\frac{V_a}{U}\left[ U^\prime\frac{\p \tilde b_y}{\p x}\mp\left( -2\frac{U}{R}\frac{\p u_x^{\text{inc}}}{\p y}+U^\prime\frac{\p u_y^{\text{inc}}}{\p x}+U^\prime\frac{\p u_x^{\text{inc}}}{\p y}\right)\right]\pm\left( - 2\frac{U}{R}\frac{\p \tilde b_x}{\p y}+U^\prime\frac{\p \tilde b_y}{\p x}+U^\prime\frac{\p \tilde b_x}{\p y}\right),
\end{split}
\end{equation}
\begin{equation}
\left(1\pm\frac{V_a}{U}\right)\frac{\partial {z}^{\mp}}{\partial t}+{\bf u\cdot\bn}z^\mp= \frac{\rho_0}{\rho}{\bf \tilde B\cdot\bn}\tilde b_z\mp{\bf \tilde B\cdot\bn}u_z\pm\tilde B_z\xi\pm\frac{z^+-z^-}{2}\left(\frac{\boldsymbol\nabla\cdot{\bf U}}{2}\mp\boldsymbol\nabla\cdot{\bf V}_a\right)\pm z^\pm\left(\frac{V_a\mp U}{R}\right).
\end{equation}
\label{2D}
\end{subequations}

\section{Analytical solution of forced Alfv\'en waves in the AEB}
\label{sing}

We consider here a shear Alfv\'en wave in a radial magnetic field ${\bf B}_0$, with solar wind speed $U$ and Alfv\'en speed $V_a$. The AEB counterpart of the equations for outwards and inwards waves are:  

\begin{equation}
\left(1+\frac{V_a}{U}\right)\frac{\partial {{ z}}^{-}}{\partial t}=- ikV_a {{ z}}^{-}+ {{ z}}^+\left(\frac{V_a-U}{R}\right)+\left(\frac{{{ z}}^+-{{ z}}^-}{2}\right)\left(\frac{\boldsymbol\nabla\cdot{\bf U}}{2}-\boldsymbol\nabla\cdot{\bf V}_a\right),
\label{aeb1b}
\end{equation}
\begin{equation}
\left(1-\frac{V_a}{U}\right)\frac{\partial {{ z}}^{+}}{\partial t}= ikV_a {{ z}}^{+}-{{ z}}^-\left(\frac{V_a+U}{R}\right)+\left(\frac{{{ z}}^--{{ z}}^+}{2}\right)\left(\frac{\boldsymbol\nabla\cdot{\bf U}}{2}+\boldsymbol\nabla\cdot{\bf V}_a\right).
\label{aeb2b}
\end{equation}

We derive below the analytical solution of eqs.~(\ref{aeb1b})--(\ref{aeb2b}) with given initial condition $z^-(x,0)=A^-(0)\exp(ik_*x)$, $z^+(x,0)=0$. We employ a two-scale approach in which the small parameter is $\epsilon=1/(k_*R_0)$, where $k_*$ is the initial wave number, $k=k_*(L_{0x}/L_x)$, and we recall that $\tau_e=R_0/U_\infty$. For the sake of simplicity we consider monochromatic waves, and according to the two-scale analysis we will look for first order solutions in $\epsilon$  of the form

\begin{equation}
z^-(x,t)=A^-\left({t/}{\tau_e}\right)\exp\left[i k_*x-i\int_0^t{\omega(t^\prime/\tau_e)}dt^\prime\right],
\label{zp}
\end{equation}
\begin{equation}
z^+(x,t)=\epsilon(z^+_f+z^+_h),
\end{equation}

where the subscript ``f" and ``h" stand for  the forced and the homogeneous part of $z^+$.

\subsection{Outward wave}
\label{out}

By substitution of (\ref{zp}) into (\ref{aeb1b}) and by introducing normalized variables

\begin{equation}
\hat t=\frac{t}{\tau_e},\quad \hat U =\frac{U}{U_\infty},\quad \hat V_a =\frac{V_a}{U_\infty}, \quad \hat\omega=\frac{\omega}{k_* U_\infty},\quad\hat\nabla=R_0\nabla,
\end{equation}

we obtain

\begin{equation}
\left(1+\frac{V_a}{U}\right)\left[  \frac{\partial A^-}{\partial \hat t} -i\frac{\hat\omega}{\epsilon}A^- \right]=-i\frac{L_0}{L}\frac{\hat V_a}{\epsilon} A^--\left(  1-\frac{V_a}{U} \right)\frac{\hat \nabla\cdot{\bf \hat U}}{4}A^-+\mathcal{O}(\epsilon^2).
\label{zm}
\end{equation}

At order $\epsilon^0$ one gets an equation for  the phase of the outwards wave

\begin{equation}
\left(1+\frac{V_a}{U} \right) i\hat\omega=i\frac{L_0}{L} \hat V_a\longrightarrow \left(1+\frac{V_a}{U} \right) \omega=kV_a\longrightarrow\omega= k_{abs}V_a, 
\end{equation}

where $k_{abs}$ is the wave vector in the absolute frame of reference, 

\begin{equation}
k_{abs}=\frac{kU}{U+V_a}=\frac{\Omega}{U+V_a}, \quad \Omega=const.
\end{equation}

At the order  $\epsilon$ one obtains an equation for the amplitude

\begin{equation}
\left(1+\frac{V_a}{U} \right)  \frac {\partial A^-} {\partial \hat t} =-\left(  1-\frac{V_a}{U} \right)\frac{\hat \nabla\cdot{\bf \hat U}}{4}A^-,
\label{ampl1}
\end{equation}

that can be integrated numerically by changing variable to $dR=Udt$, and has solution, as expected,

\begin{equation}
A^-(t)=C_0\frac{\sqrt{UV_a}}{U+V_a},
\end{equation}
where $C_0$ is determined from the initial condition.

\subsection{Inward forced wave}
Assuming that there are no inwards waves at $t=0$, we look for a solution to eq.~(\ref{aeb2b}) that in general is given by a forced wave $z^+_f$, corresponding to the particular solution of eq.~(\ref{aeb2b}), plus a reflected wave $z^+_h$, the latter being the solution of the homogeneous part of eq.~(\ref{aeb2b}). The forced wave is of the form

\begin{equation}
z^+_f(x,t)=\epsilon A_f^+(t/\tau_e)\exp\left[ik_* x-i\int_0^t \Omega \frac{V_a(t^\prime/\tau_e)}{U(t^\prime/\tau_e)+V_a(t^\prime/\tau_e)} dt^\prime\right].
\label{forcedeq}
\end{equation}

By inserting~(\ref{forcedeq}) in eq.~(\ref{aeb2b}) and with the same ordering procedure as above, one gets to first order in $\epsilon$ an algebraic equation for $A^+$ (remember that $k=k_* L_{0x}/L_x$):

\begin{equation}
-i\left[ \left(  1-\frac{V_a}{U} \right)k V_a \frac{U }{U+V_a} +V_ak\right]  A_f^+=A^-\left[  \frac{\nabla\cdot{\bf U}}{4}\left(1+\frac{V_a}{U}\right) -\left( \frac{U+V_a}{R} \right)\right].
\label{forced}
\end{equation}

The amplitude of $z_f^+$ is therefore

\begin{equation}
A_f^+(t)=-iA^-(t)\left(\frac{1}{2R}  + \frac{U^\prime}{4U}\right) \frac{(U + V_a)^2}{2\Omega V_a}.
\end{equation}

The homogeneous solution $z_h^+$ is necessary in order to impose the initial condition, and must satisfy the following equation,

\begin{equation}
\left(  1-\frac{V_a}{U} \right)\frac{\partial z^+_h}{\partial t}=ikV_a z^+_h,
\label{homo}
\end{equation}

and hence, to leading order in $\epsilon$,

 \begin{equation}
z^+_h(x,t)=A^+_h\exp\left[ik_*x+i\int^t \frac{\Omega V_a}{U-V_a}dt^\prime-i\phi\right].
\end{equation}

By imposing the initial condition $z^+_h(x,0)+z^+_f(x,0)=0$ we obtain a condition on $\phi$ and $A^+_h$,

 \begin{equation}
A^+_h\,e^{-i\phi}+A^+_f(0)=0\quad\Rightarrow\quad e^{-i\phi}=-1, \, A^+_h=A_f^+(0).
\end{equation}

The complete solution of the $z^+$ mode is,

 \begin{equation}
z^+(t)=A^+_f(t)e^{ik_*x-i\Omega\int^t\frac{ V_a}{U+V_a}dt^\prime}\left(1-\frac{A^+_f(0)}{A^+_f(t)}e^{i\Omega\int^t\frac{2V_aU}{U^2-V_a^2}dt^\prime}   \right),
\end{equation}

from which it can be seen that the total amplitude of $z^+$ displays beating oscillations at a frequency corresponding to the phase difference between the forced and the reflected waves:

 \begin{equation}
|z^+(t)|^2=A_f^2(t)+A_f(0)^2-2A_f(t)A_f(0)\cos\Phi,\quad\Phi=\Omega\int^t\frac{2V_aU}{U^2-V_a^2}dt^\prime.
\end{equation}


\end{document}